\documentclass[floats,floatfix,amssymb,prd,twocolumn,superscriptaddress,nofootinbib,preprintnumbers]{revtex4-1}

\usepackage{subcaption,ragged2e}
\DeclareCaptionJustification{justified}{\justifying}
\usepackage{amssymb,amsmath,verbatim,mathtools,needspace,enumitem,etoolbox,graphicx,microtype,afterpage,bm}

\usepackage{tikz}
\usetikzlibrary{shapes.misc, positioning, arrows.meta}
\usepackage{framed}

\usepackage{hyperref}
\hypersetup{colorlinks, citecolor=bluscuro, linkcolor=black, urlcolor=bluscuro}
\definecolor{rossos}{cmyk}{0,1,1,0.55}
\definecolor{bluscuro}{rgb}{0.15, 0.2, .85}
\definecolor{bluchiaro}{cmyk}{1,.3,0.,0.1}
\definecolor{ForestGreen}{rgb}{0.13, 0.55, 0.13}
\definecolor{TLGreen}{RGB}{50, 164, 49}
\definecolor{TLOrange}{RGB}{231,180,22}
\definecolor{TLRed}{RGB}{204,50,50}

\usepackage{multirow,pifont,lmodern,float,tabularx,booktabs}
\usepackage[all]{hypcap}
\usepackage[T1]{fontenc}
\usepackage[utf8]{inputenc}

\renewcommand{\arraystretch}{1.4}

\allowdisplaybreaks

\newcommand{\be}{\begin{equation}}
\newcommand{\ee}{\end{equation}}
\renewcommand{\d}{{\rm d}}

\newcommand{\cern}{
CERN, Theoretical Physics Department,
Esplanade des Particules 1, Geneva 1211, Switzerland}

\newcommand{\IAP}{Institut d'Astrophysique de Paris, UMR 7095 du CNRS et de Sorbonne Universit\'e,\\ 98 bis bd Arago, 75014 Paris, France}

\begin{document}

\title{Ultra-Slow-Roll Inflation on the Lattice I:\\ Backreaction and Nonlinear Effects}
\author{Angelo Caravano}
\email{caravano@iap.fr}
\affiliation{\IAP}

\author{Gabriele Franciolini}
\email{gabriele.franciolini@cern.ch}
\affiliation{\cern} 

\author{Sébastien Renaux-Petel}
\email{renaux@iap.fr}
\affiliation{\IAP}

\date{\today}

\begin{abstract}
\noindent
Violating the slow-roll regime during the final stages of inflation can significantly enhance curvature perturbations, a scenario often invoked in models producing primordial black holes and small-scale scalar induced gravitational waves. When perturbations are enhanced, one approaches the regime in which tree-level computations are insufficient, and nonlinear corrections may become relevant. In this work, we conduct lattice simulations of ultra-slow-roll (USR) dynamics to investigate the significance of nonlinear effects, both in terms of backreaction on the background and in the evolution of perturbations. Our systematic study of various USR potentials reveals that nonlinear corrections are significant when the tree-level curvature power spectrum peaks at $\mathcal{P}^{\rm max}_{\zeta} = {\cal O}(10^{-3})-{\cal O}(10^{-2})$, with {5\%$-$20\%} corrections. Larger enhancements yield even greater differences. We find a simple universal relation between simulation and tree-level quantities $\dot\phi = \dot\phi_{\rm tree}\left(1+\sqrt{\mathcal{P}^{\rm max}_{\zeta,\rm tree}}\right)$ at the end of the USR phase, which is valid in all cases we consider. Additionally, we explore how nonlinear interactions during the USR phase affect the clustering and non-Gaussianity of scalar fluctuations, crucial for understanding the phenomenological consequences of USR, such as scalar-induced gravitational waves and primordial black holes. Our findings demonstrate the necessity of going beyond leading order perturbation theory results, through higher-order or non-perturbative computations, to make robust predictions for inflation models exhibiting a USR phase.
\end{abstract}

\preprint{CERN-TH-2024-181}

\maketitle

{
\setcounter{tocdepth}{1}
  \hypersetup{linkcolor=black}
  \tableofcontents
}
\hypersetup{linkcolor=bluscuro}

\section{Introduction}\label{intro}

The inflationary paradigm currently represents the leading explanation for the early Universe's evolution, successfully predicting a nearly flat spatial geometry and the properties of primordial fluctuations observed in the cosmic microwave background (CMB) and large-scale structure. These fluctuations are nearly scale-invariant, Gaussian, adiabatic, and dominated by a growing scalar mode, resulting in a Universe that is statistically homogeneous and isotropic~\cite{Akrami:2018odb, Starobinsky:1980te, Guth:1980zm, Linde:1981mu, Albrecht:1982wi}. In the simplest inflationary models, a scalar field called the inflaton moves slowly down its potential, balanced by Hubble friction, resulting in a ``slow-roll'' (SR) phase.

However, in certain models, the inflaton potential features a flat region or shallow minimum where the deceleration becomes dominant, and the field's velocity can decrease exponentially, in what is called an ``ultra-slow-roll'' (USR) phase \cite{Ivanov:1994pa,Kinney:1997ne,Inoue:2001zt,Kinney:2005vj,Martin:2012pe,Motohashi:2017kbs}. Perturbations generated during USR are far from scale-invariant, displaying a large amplification. These perturbations may lead to interesting phenomenological consequences, such as the emission of scalar-induced gravitational waves (SIGW) \cite{Tomita:1975kj, Matarrese:1993zf, Acquaviva:2002ud, Mollerach:2003nq, Ananda:2006af, Baumann:2007zm, Domenech:2021ztg}, and/or seed primordial black hole (PBH) formation
\cite{Zeldovich:1967lct,Hawking:1971ei,Carr:1974nx,Carr:1975qj,Chapline:1975ojl}. 
PBHs are a long-standing candidate for dark matter, potentially accounting for all dark matter or acting as seeds for supermassive black holes, see Refs.~\cite{Carr:2020gox, Green:2020jor} for recent reviews. The blooming field of GW astronomy will provide unprecedented observations of small-scale GWs, being able to constrain the early universe (see \cite{Achucarro:2022qrl,LISACosmologyWorkingGroup:2022jok,LISACosmologyWorkingGroup:2023njw} for recent reviews).

A very relevant question is whether the dynamics of USR are consistent with perturbativity (see Ref.~\cite{Kristiano:2024ngc} for a recent review). From a technical standpoint, the dimensionless power spectrum of curvature perturbations, $\mathcal{P}_\zeta(k)$, is usually calculated within linear, i.e. tree-level, perturbation theory. However, as linear perturbations can become large (often reaching $\sqrt{\mathcal{P}_\zeta(k)} \simeq 0.1$), curvature perturbations are expected to be sensitive to nonlinear interactions. 
These nonlinear effects have various consequences.
First, they introduce nonzero higher-order cumulants in addition to the variance, with important implications for the PBH formation, see e.g.~\cite{Franciolini:2018vbk,Taoso:2021uvl}. Second, through loops, they modify the variance itself compared to the result obtained from the free theory.
Lastly, they lead to a backreaction on the background dynamics. 
In this paper we investigate the relevance of these effects, considering motivated scenarios built with the reverse engineering approach proposed in Ref.~\cite{Franciolini:2022pav}, by performing numerical lattice simulations able to capture the dynamics of the inflaton non-perturbatively.

This paper is organized as follows. We describe the toy model inflaton potential we build to test various USR scenarios in Sec.~\ref{sec:setup}, where we also discuss the tree-level predictions of standard perturbation theory (SPT). 
In Sec.~\ref{sec:lattice}, we describe the lattice simulations performed to test the relevance of nonlinear corrections, both on the background evolution as well as on the spectrum of curvature perturbations. We then compare the numerical results to the tree-level predictions, before concluding in Sec.~\ref{sec:conclusions} with some outlook directions left for future work. 
We set $M_{\textrm{Pl}}^2 = 1/(8\pi G) = 1$ throughout this work.

\section{Reverse engineered USR toy models}
\label{sec:setup}

We describe here how we construct the inflationary potentials we consider in the lattice simulations. We stress again that these scenarios are explicitly built in order to explore specific properties of transient USR phases, and should be considered as toy models. While they can be realised in first principle models of inflation (see e.g. \cite{Ballesteros:2017fsr,Karam:2022nym,Franciolini:2022pav}), we do not attempt to perform this connection here, and leave this task for future work. 
For simplicity, we set ourselves in the three phase model of inflation, in which the dynamics is described by three regimes: SR - USR - SR.
We assume a relatively fast transition between each regime, although always considering realistic transitions away from the instantaneous transition limit, lasting $\lesssim 1$ e-folds. After having fixed the parameters controlling the inflationary dynamics, we reconstruct the inflationary potential that realise such scenarios at tree-level.  
In the following, we adopt the procedure laid out in Ref.~\cite{Franciolini:2022pav}.

\subsection{Inflationary background}\label{sec:inflback}
The inflationary background can be
described by modeling the evolution of the Hubble rate $H \equiv \dot a/a$, where $a$ is the scale factor during inflation.
This is dictated by dynamical equations relating $H$  
to the Hubble parameters, which are 
\begin{align}\label{eq:HubbleParameters}
\epsilon\equiv -\frac{\dot{H}}{H^2}\,,~~~~~~~~~
\eta \equiv -\frac{\ddot{H}}{2H\dot{H}} = \epsilon - \frac{1}{2}\frac{d\log\epsilon}{dN}\,,
\end{align} 
where $\dot{H} = dH/dt$ is the cosmic-time derivative of $H$ while
$N$, defined as $dN = Hdt$, is the number of $e$-folds.\footnote{
Notice that a different definition of the Hubble parameter $\eta$ is sometimes used, and defined as 
$\epsilon_2 \equiv \dot \epsilon / H \epsilon = -2 (\eta-\epsilon)$.}

\begin{figure}[t!]
\centering
\includegraphics[width=0.5\textwidth]{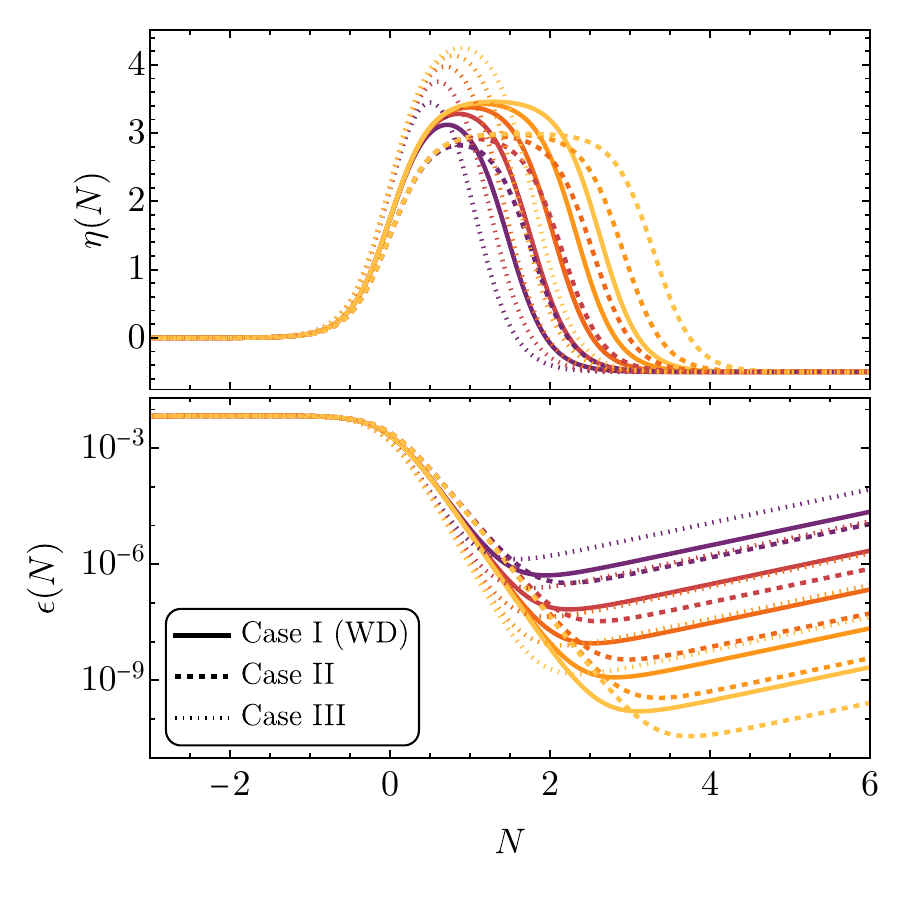}
\caption{ 
Evolution of $\eta (N)$ (top panel) and $\epsilon(N)$ (bottom panel) as a function of number of $e$-folds considered in this work. Solid, dashed and dotted lines indicate different cases, corresponding to the approximate Wands duality (WD), repulsive and attractive case, respectively. See the main text for more details. Different colors (from purple to yellow) correspond to different USR durations $N_{\rm  end} - N_{\rm  in}$, leading to a tree-level curvature power spectrum at its maximum ranging from ${\cal P}_\zeta^{\rm  max} =10^{-4}$ to ${\cal P}^{\rm  max}_\zeta =1$. In all cases, we fixed $N_{\rm  in } = 0$.
}\label{fig:dyn}
\end{figure}

We now introduce a simple semi-analytical model\,\cite{Byrnes:2018txb,Taoso:2021uvl,Franciolini:2022pav} to describe the evolution of $\eta$ through the SR-USR-SR transitions. 
We define the hyperbolic tangent parametrization
\begin{align}
&\eta(N) = \frac{1}{2}\left[
-\eta_{\rm II}+ \eta_{\rm II}\tanh\left(\frac{N-N_{\rm in}}{\delta N}\right)
\right] 
\nonumber \\
&+ \frac{1}{2}\left[
\eta_{\rm II} + \eta_{\rm III} + (\eta_{\rm III}-\eta_{\rm II})\tanh\left(\frac{N-N_{\rm end}}{\delta N}\right)
\right]\,,\label{eq:DynEta}
\end{align}
where the parameter $\delta N$ controls the width of the two transitions at $N_{\rm in}$ and $N_{\rm end}$. We assume that the second Hubble parameter $\eta$ is negligible in the initial slow-roll phase $\eta_{\rm I} \simeq 0$, consistent with the CMB requirement in the earlier stage of inflation.
Once the parameters in Eq.~\eqref{eq:DynEta} are fixed, one can solve for the evolution of $\epsilon$ and $H$ as a function of number of $e$-folds using \eqref{eq:HubbleParameters}. In all cases, we arbitrarily set $N_{\rm  in} = 0$. 
In Fig.~\ref{fig:dyn}, we show the evolution of $\epsilon$ and $\eta$ for the scenarios considered in this work, which can be divided into three cases, depending on the value of $\eta$ during USR:
{\it i)} the approximate Wands duality (case I);
{\it ii)} the repulsive (case II);
{\it iii)}  the attractive (case III). The choice of these names will become clear in the following.

In Tab.~\ref{tab:usr_cases} we report parameters we adopt in the various cases. 
We arbitrarily fix $\delta N = 0.5$, in order to have relatively smooth transitions in and out of the USR phase. 
Additionally, we choose values of $\eta_{\rm II}$ that are in all cases  larger than $\simeq 3/2$, as required to enter in the USR regime. Finally, we set  $\eta_{\rm III} = -0.5$, as a negative value is required in order to end inflation by raising $\epsilon$ again to reach ${\cal O}(1)$.\footnote{For simplicity, we use the terminology ``slow-roll'' for the third phase, by contrast with the preceding USR phase, despite the relatively large value of $\eta_{\rm III}$.}
In case I, we approximately realise a scenario that respects the Wands duality (WD) \cite{Wands:1998yp}, stating that the same tree-level spectrum of perturbations is obtained during phases characterised by a constant $\eta$ or $3-\eta$.
In case I, this property is respected between phases II and III, and therefore it 
leads to a tree-level curvature power spectrum featuring a single smooth peak that can be fitted with a broken power-law. 
This happens because the spectrum of modes exiting the Hubble scale during USR follows $k^{2(3-\eta_{\rm II})}\sim 1/k$, while $k^{2\eta_{\rm III}}\sim 1/k$ in the subsequent phase \cite{Karam:2022nym,Franciolini:2022pav}.
This property can also be derived by inspecting the effective mass of scalar perturbations when $\epsilon$ is negligible. We will come back to this point in Sec.~\ref{sec:wandsdual}. As we will see, the nearly constant effective mass term will also be accompanied by negligible higher-order derivatives of the potential, rendering the theory nearly free from self-interactions at these times. 
In cases II and III, $\eta_{\rm II}$ is taken to be smaller or larger than the WD value, leading to different phenomenology for the scalar perturbations.

{
\renewcommand{\arraystretch}{1.4}
\setlength{\tabcolsep}{4pt}
\begin{table}
\caption{ Parameters adopted in the various scenarios considered in this work. The value of $\Delta N$ is given as a function of the maximum ${\cal P}_\zeta$ obtained with the tree-level computation. }
\begin{tabularx}{1 \columnwidth}{|X|c|c|c|}
\hline
\hline
& $\eta_{\rm II}$ & $\eta_{\rm III}$ & $\Delta N \equiv N_{\rm  end} - N_{\rm  in}$ 
\\
\hline
Case I (Wands duality) & $3.5$ & $-0.5$ &
$2.6+ 
0.29 \log_{10} {\cal P}_\zeta^{\rm max} 
$
\\
\hline
Case II (repulsive) & $3$& $-0.5$ &
$3.3+ 
0.38 
\log_{10} {\cal P}_\zeta^{\rm max} 
$
\\
\hline
Case III (attractive)  &$4.5$ & $-0.5$ & 
$1.8+ 
0.19 \log_{10} {\cal P}_\zeta^{\rm max} 
$
\\
\hline
\hline
\end{tabularx}
\label{tab:usr_cases}
\end{table}
}

\begin{figure*}[t!]
\centering
\includegraphics[width=0.32\textwidth]{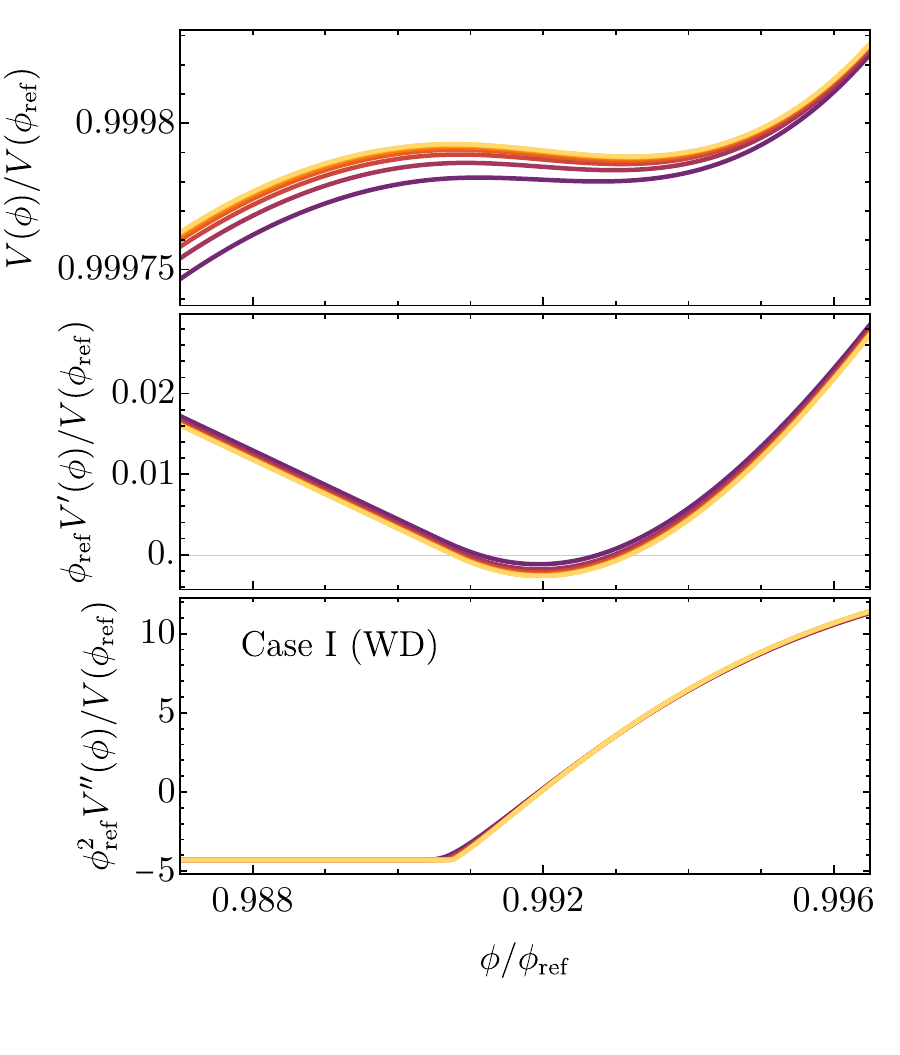}
\includegraphics[width=0.32\textwidth]{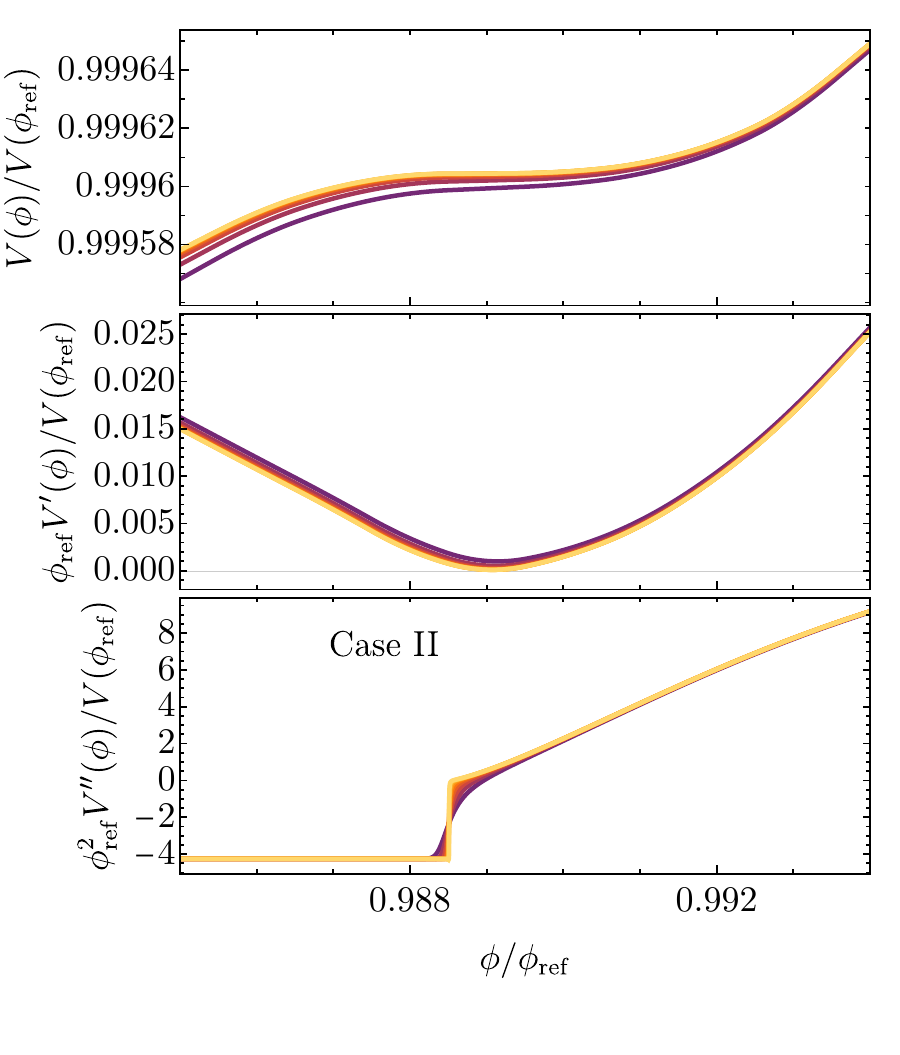}
\includegraphics[width=0.32\textwidth]{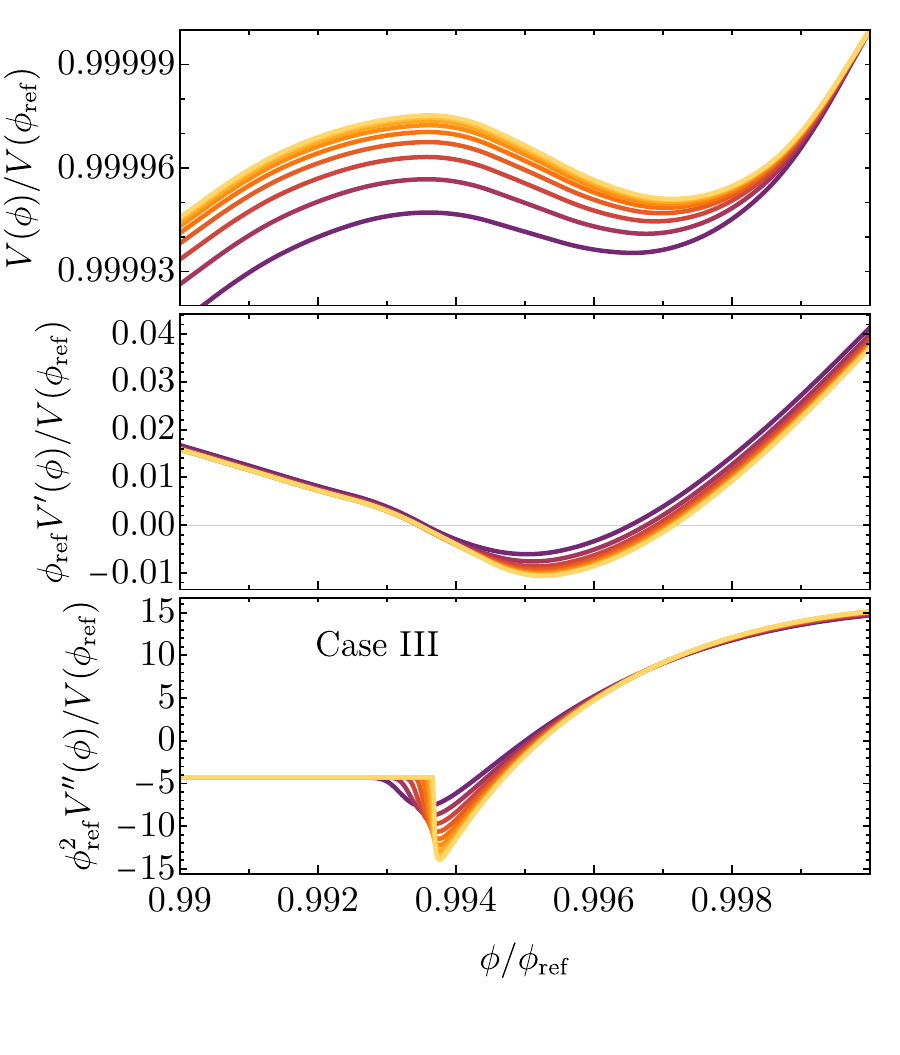}
\caption{ 
Reconstructed inflationary potentials built with the reverse engineering approach and choosing the parameters reported in Tab.~\ref{tab:usr_cases}.
From top to bottom, we show the potential  and its first and second derivatives with respect to the inflaton field $\phi$, normalised to their quantities at $N_{\rm in}$, i.e. $\phi_{\rm ref}$ and $V_{\rm ref}$.
{\it Left panel:} case I (WD);
{\it Center panel:} case II (repulsive);
{\it Right panel:} case III (attractive). 
}\label{fig:dyn_pot}
\end{figure*}

\subsection{Inflaton potential}

Once the Hubble parameters are known, one can compute the inflationary potential by means of \cite{Byrnes:2018txb}
\begin{subequations}
\begin{align}
V(N) & = V(N_{\rm CMB})\exp\left\{
   -2\int_{N_{\rm CMB}}^{N}dN^{\prime}\left[\frac{\epsilon(3-\eta)}{3-\epsilon}\right]
   \right\}\,,
   \label{eq:recPot1}
\\
\phi(N) & = \phi(N_{\rm CMB}) \pm \int_{N_{\rm CMB}}^N dN^{\prime}\sqrt{2\epsilon}\,, 
\label{eq:recPot2}
\end{align}
\end{subequations}
where in the second equation we consider the minus sign having in mind a large-field model in which the field value decreases as inflation proceeds. 
Combining $V(N)$ and $\phi(N)$, we reconstruct the 
profile $V(\phi)$ of the inflationary potential in field space. 
In all cases, we arbitrarily choose $\phi(N_{\rm CMB}) = 3.5$ and, by fixing the amplitude of the power spectrum at large scales to be compatible with CMB data, i.e. ${\cal P}_\zeta = 2.1 \cdot 10^{-9}$, we find 
$V(N_{\rm CMB}) = 3.2
\cdot 10^{-9}$.
The initial condition satisfying the SR attractor at $N_{\rm CMB}$ require $\phi(N_{\rm CMB}) = 3.5$ and 
$\d \phi(N_{\rm CMB}) / \d N = - 0.11$.
Eq.\,\eqref{eq:recPot1} shows the convenience of modeling the inflationary dynamics directly at the level of $\eta$ instead of $V(\phi)$. 
This is because the Hubble parameters enters at the exponent of the definition of $V(N)$, and thus allow for a much finer control on power spectral features 
when performing the reverse engineering procedure.

In Fig.~\ref{fig:dyn_pot}, we show the reconstructed potential for the various scenarios considered in this work, zooming in the shallower region of the potential that controls the USR dynamics. 
We also show the first and second derivatives of the inflaton potential, as they will be relevant for the subsequent discussion. 
It is interesting to notice that, in cases II and III, sharp changes of the second derivative of the potential are obtained near the end of the feature, corresponding to the expected end of the USR phase. This will lead to large higher-order derivatives of the potential, crucially controlling the coupling of the theory. 
Notice that the potentials of case II are monotonic. Cases I and III present instead a local minimum followed by a local maximum (in the direction of decreasing $\phi$). However, the minimum is not deep enough to lead to a trapping of the inflaton \cite{Atal:2019cdz,Atal:2019erb,Inomata:2021tpx,Escriva:2023uko,Caravano:2024tlp} (contrary to what happens in other contexts, see e.g. \cite{Caravano:2024tlp} in a model with oscillations in the potential \cite{Inomata:2022yte}), with the exception of the most extreme situation of case III with $\mathcal{P}^{\rm max}_{\zeta,\rm tree}= 1$, as we will discuss below.

Using the reconstructed potential $V(\phi)$, 
one can also solve the inflaton equation of motion 
\begin{align}\label{eq:EoM}
\frac{d^2\phi}{dN^2} + \left[3 - \frac{1}{2}\left(\frac{d\phi}{dN}\right)^2\right]
\left[\frac{d\phi}{dN} + \frac{V_{,\phi}(\phi)}{V(\phi)}
\right] = 0\,,
\end{align}
and, in turn, compute the time evolution of the Hubble parameters in Eq.\,(\ref{eq:HubbleParameters}) 
and the Hubble rate by means of the relations
\begin{subequations}
\begin{align}
H^2 & =\frac{V(\phi)}{3-\epsilon},
 \\
\epsilon &= \frac{1}{2}\left(\frac{d\phi}{dN}\right)^2\,,
\\
\eta &= 
3 - \frac{V_{,\phi}(\phi)[- 6 + (d\phi/dN)^2]}{2V(\phi)(d\phi/dN)}
\,.
\end{align}
\end{subequations}
These solutions can be compared to the input of the model to test for the numerical precision of the reconstruction of the potential.

\subsection{Tree-level curvature power spectrum}

Once the background evolution is defined, we can calculate the spectrum of gauge-invariant comoving curvature perturbations generated during inflation. As long as the slow-roll approximation holds, this spectrum is given by
$
{\cal P}_{\zeta}(k) = {H^2}/{8\pi^2 \epsilon}
$
where the Hubble parameters are evaluated at the Hubble crossing of mode $k$. This hints toward the general expectation of exponentially enhanced perturbations during the USR phase, where $\epsilon$ decays exponentially.

We compute the tree-level prediction without approximation by solving the Sasaki-Mukhanov (SM) equation \cite{Sasaki:1986hm,Mukhanov:1988jd}
\begin{align}\label{eq:SM-full}
\frac{\d^2 u_k}{\d N^2}+(1-\epsilon) \frac{\d u_k}{\d N}+ \left[ \left(\frac{k}{a H} \right)^2-2+m_\textrm{eff}^2/H^2\right]u_k= 0\,,
\end{align}
 where the effective mass reads
\begin{align}
m_\textrm{eff}^2/H^2=\eta(3-\eta)+\d\eta/\d N+3 \epsilon \eta-2 \epsilon -2 \epsilon^2.
\label{eq:full-mass-term}
\end{align}
We solve the SM equation with sub-Hubble Bunch-Davies initial conditions for modes deep within the Hubble sphere, i.e. when $N$ is smaller than the Hubble crossing epoch $N_k$ such that $k = a(N_k) H(N_k)$. This is implemented  on the SM variable as
\begin{align}
u_k = \frac{1}{\sqrt{2 k}}, 
\qquad
\frac{du_k }{dN}  = -  \frac{k}{\sqrt{2} a(N)H(N)}, 
\end{align}
for $N \ll N_k$, where we choose the phase of $u_k$ to be real initially without loss of generality.

We then compute the power spectrum ${\cal P}_{\zeta}(k)$ of the gauge-invariant comoving curvature perturbation $\zeta$
\begin{align}\label{eq:PS}
{\cal P}_\zeta(k) = \frac{k^3}{2\pi^2}\left|\frac{u_k(N_f)}{z(N_f)}\right|^2\,,
\end{align}
where $z(N) = a(N) {\d\phi(N)}/{\d N}$. 
The power spectrum ${\cal P}_\zeta(k)$ is time-independent since it is evaluated after the modes freeze on sufficiently super-Hubble scales \cite{Lyth:2004gb}.

\begin{figure*}[t!]
\centering
\includegraphics[width=0.49\textwidth]{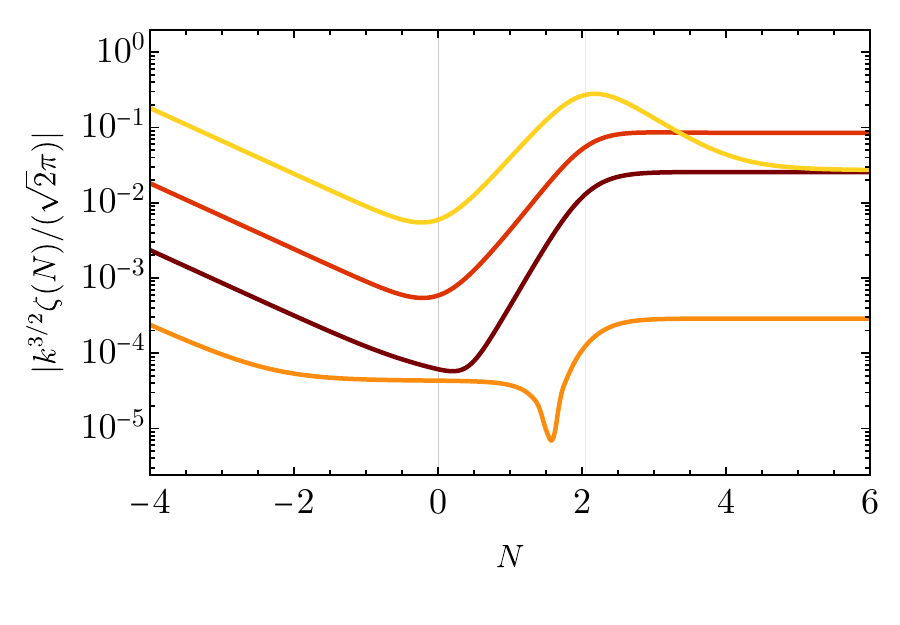}
\includegraphics[width=0.49\textwidth]{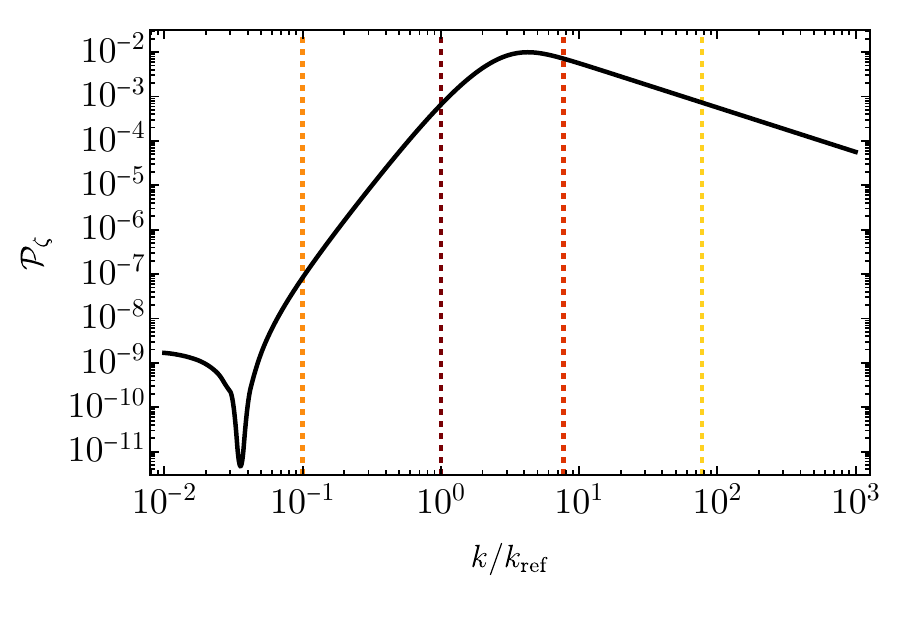}
\caption{Illustration of the tree-level predictions of Standard Perturbation Theory, considering for definiteness Case I and $\Delta N = 2$ (corresponding to a very good approximation with $\mathcal{P}^{\rm max}_{\zeta,\rm tree}=10^{-2}$).
{\it Left panel:}
Curvature perturbation mode evolution as a function of number of $e$-folds. At early times, modes evolve from the Bunch-Davies vacuum. 
The orange line indicates the evolution of a mode crossing the Hubble sphere before the USR phase. Indeed, we see it freezing out, and then experiencing a change of behavior for the decaying mode, until the USR phase ends again. 
Brown and red lines correspond to modes crossing the Hubble scale at the onset and end of USR, respectively. The yellow line corresponds to a mode crossing the Hubble sphere after USR has ended, but sufficiently close to it to be affected.
{\it Right panel:}
Curvature power spectrum near the peak. Vertical dashed lines indicate the scales with the same color code adopted in the left panel. For reference, the lattice simulations stop at around $N=5$, after the relevant modes have frozen. 
}\label{fig:pzeta}
\end{figure*}

To illustrate the typical dynamics, Fig.~\ref{fig:pzeta} shows the evolution of curvature perturbations as a function of the number of $e$-folds for Case I with $\Delta N = 2$ (corresponding to a very good approximation with $\mathcal{P}^{\rm max}_{\zeta,\rm tree}=10^{-2}$).
In the left panel, we show modes crossing the Hubble sphere before, during and right after the USR phase. While each mode experiences a growth due to the decaying mode changing behavior during USR, they all freeze out soon after USR ends and the subsequent SR evolution begins. The lattice simulations track the evolution of the system until the relevant modes have frozen.
In the right panel, we indicate where these modes sit in the curvature power spectrum at the end of inflation. We normalize the x-axis with respect to $k_{\rm ref}$, which is the mode crossing the Hubble scale at $N_{\rm in}$, i.e. $k_{\rm ref} = a(N_{\rm in}) H(N_{\rm in})$.

\subsubsection{On the irrelevance of gravitational backreaction}
\label{sec:grav_back}

We can inspect the relevance of the gravitational backreaction by focusing on the SM equation \eqref{eq:SM-full}. 
By neglecting gravitational backreaction, i.e.~neglecting the lapse and shift perturbations in the spatially flat gauge, one obtains a different mass term, namely Eq.~\eqref{eq:full-mass-term} subtracting $4 \epsilon \eta-6 \epsilon -2 \epsilon^2$. In the situations of interest we have $\epsilon \ll 1$, which shows that gravitational backreaction can be safely neglected as long as $|3-\eta| \gg \epsilon$. 
This simply corresponds to the decoupling limit (see e.g. \cite{Creminelli:2024cge}) in the effective field theory of inflation \cite{Cheung:2007st}.
This condition is satisfied in our setups and hence our lattice simulations, which do not take into account metric fluctuations, should agree with tree-level results if standard perturbation theory holds.\footnote{Note that, if necessary, metric perturbation can be perturbatively taken into account in lattice simulations, as recently done in \cite{Caravano:2024xsb}.}
We have confirmed this explicitly by checking the relative difference between the effective mass terms computed in the two cases. Due to the exponential decay of $\epsilon$ during the USR phase $\epsilon \sim \exp\left ( - 2 \eta_{\rm II} N \right) $, the backreaction induced terms quickly become subdominant, while $m_{\rm eff}^2/H^2$ remains ${\cal O}(1)$.

\subsubsection{Action for perturbations with and without Wands duality}\label{sec:wandsdual}

As shown for the inflaton mass term above, in the regime where $\epsilon$ is small,
one falls in the decoupling limit where interactions that arise from the metric fluctuations are suppressed (see also \cite{Ballesteros:2024zdp}).
The action for inflaton perturbations in the flat gauge can then be written as
\begin{align}	\label{eq:pert_act}
S = {\int}
\d \tau\, \d^3 x
 a^2
 \, \left[
 \frac{1}{2}
\left(\partial_\tau\delta\phi\right)^2 - 
\frac{1}{2}\left(\partial_i\delta\phi\right)^2 -a^2 \sum_{n \geq 2} \frac{V_n\, \delta\phi\,^n}{n!}\right]
\end{align}
and we defined $V_n = \d^n V / \d \phi^n$. 
Consistently neglecting $\epsilon$-suppressed terms, one finds 
(e.g. \cite{Franciolini:2023agm,Ballesteros:2024zdp})
\begin{subequations}
\begin{align} 
\label{eq:couplings}
a^2 V_2 &= -(aH)^2 (\nu^2-9/4)\,, 
\\
a^2V_3 &= - \textrm{sign}(\dot\phi)\dfrac{aH (\nu^2)'}{\sqrt{2\epsilon}}\,, 
\label{eq:cubiccoupling}
\\
a^2V_4 &= -\dfrac{1}{2\epsilon} \left[ (\nu^2)'' - aH (\nu^2)'\left(1- \eta \right)  \right]\,,
\end{align}
\end{subequations}
where primes denote derivatives with respect to conformal time, and we introduced
\begin{align}\label{eq:nu2}
\nu^2 \equiv \frac{9}{4}
-  \left[
\eta \left (3\, - \eta\right) + \frac{\eta'}{aH}\right]\,.
\end{align}
Note that in the decoupling limit that we consider, one simply has $V_2/H^2 \simeq m^2_{\textrm{eff}}/H^2 \simeq 9/4- \nu^2$. Higher-order derivatives simply follow from $\dot{V}_{n}=\dot{\phi} V_{n+1}$.\footnote{Notice that \cite{Ballesteros:2024zdp} implicitly assumed a positive inflaton velocity and hence does not have $\textrm{sign}( \dot \phi)$ in the corresponding Eq.~\eqref{eq:cubiccoupling}.}

\begin{figure}
\centering
\includegraphics[width=0.49\textwidth]{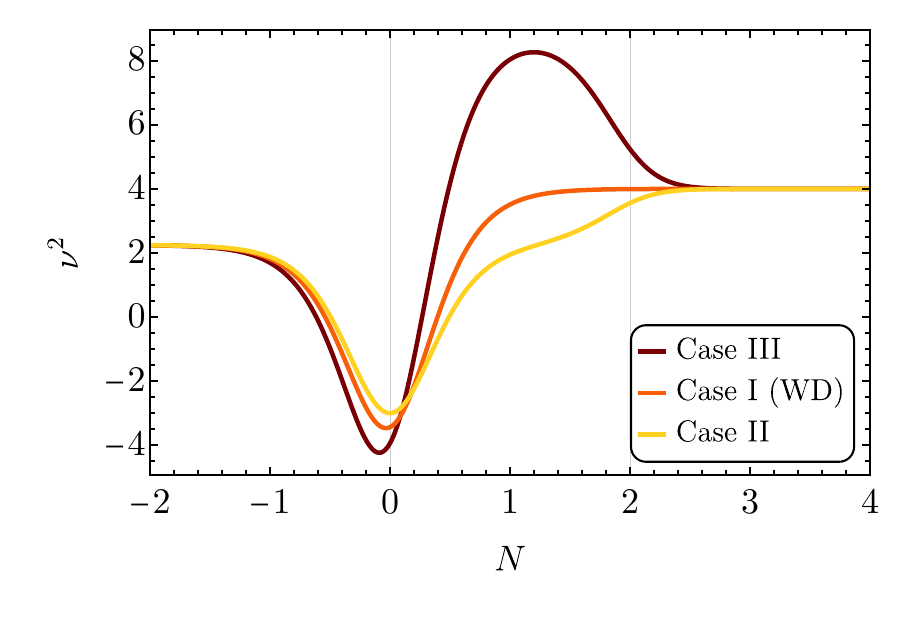}
\includegraphics[width=0.49\textwidth]{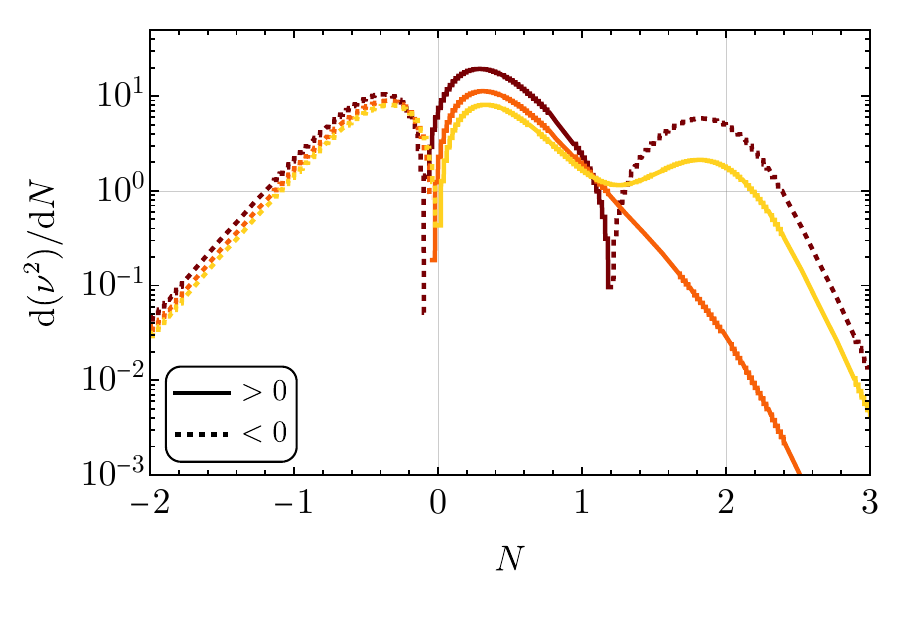}
\centering
\caption{
{\it Top panel:} Evolution of $\nu^2$ as a function of number of $e$-folds across SR-USR-SR transitions in the three cases considered in this work, with $\Delta N = 2$ in each case.
While all cases presents a sizeable evolution of $\nu^2$ at the onset of USR, in case I, $\nu^2 $ is nearly constant from deep in the USR phase. 
{\it Bottom panel:} first derivative of $\nu^2$ with respect to number of efolds. By construction, at the end of USR around $N=2$, it is ${\cal O}(1)$ and positive (resp. negative) in case II (resp. case III), which makes the former a ``repulsive'' scenario while the latter can be considered as ``attractive''. See the main text for an explanation of the terminology adopted here. By contrast, in case I with approximate Wands duality, $(\nu^2)'$ is two orders of magnitude smaller than the other cases at $N=2$, corresponding to an approximately free theory.}\label{fig:dnu}
\end{figure}

An interesting class of scenarios are the ones that respect the Wands duality \cite{Wands:1998yp}, which relates different background evolutions to the same Mukhanov-Sasaki equation. 
As discussed in Sec.~\ref{sec:inflback}, for a constant $\eta$, the simplest example of Wands duality relates USR with $\eta  = 3$ and SR with $\eta =0$.
This can be extended beyond this simple example, to scenarios with a different evolution of $\eta$ that however still approximately preserves the effective mass, which means nearly constant $\nu^2$ in Eq.~\eqref{eq:nu2}. As a consequence, in these scenarios, all the higher-order derivatives of the inflaton potential, which are proportional to time derivatives of $\nu$, are suppressed. Therefore, transient USR scenarios that approximately respect the Wands duality become almost free theories \cite{Kristiano:2024vst}.

We realise this scenario, albeit approximately, in case I. We consequently fix $\eta$ in phases II and III, while considering a symmetric transition between them such that it retains a nearly constant $\nu$ after the onset of USR, and crucially during the transition between USR and the subsequent SR.
Therefore, we only expect a backreaction on the background evolution and no effect on the spectrum of $\delta \phi$ perturbations in case I.
Notice that this property is independent from the assumed duration of USR.

We show the evolution of $\nu^2$ as a function of number of $e$-folds in Fig.~\ref{fig:dnu}. We can notice that in all cases, 
$\nu^2 $ transitions from the value $\approx 9/4$ to negative during the onset of USR, while relaxes back to $\approx 4$ afterward. The final value is obtained by inserting $\eta_{\rm III} = -0.5$ into Eq.~\eqref{eq:nu2}.
Case I approaches a nearly constant value already during USR, suppressing interactions during the last stages, including at the USR-SR transition. In the other cases, the time evolution remains sizeable, with $\nu^2$ increasing in case II, and decreasing in case III, at the critical period corresponding to the end of the USR phase (around $N \sim 2$). This corresponds to, respectively, a positive and negative cubic coupling $V_3$ in \eqref{eq:cubiccoupling}, conventionally referred to as repulsive and attractive interactions.
Therefore, in the following, we will refer to case II as the repulsive case and case III as the attractive case. We will comment more precisely on the concrete meaning of these names when analyzing snapshots of lattice simulations of the three cases in Fig.~\ref{fig:snapshots}.

\section{Lattice simulations}\label{sec:lattice}

\subsection{The method}
To go beyond the perturbative description, we use the lattice simulations of inflation recently developed in \cite{Caravano:2021pgc,Caravano:2022epk,Caravano:2022yyv}. In the following, we give a brief description of the lattice methodology and key conceptual aspects, while we refer to \cite{Caravano:2021pgc,Caravano:2022yyv} for technical details regarding the numerical calculation.
\subsubsection{Equations of motion}
The simulation allows us to solve the equation of motion for the inflaton field $\phi(\vec x,t)$
\begin{align}
\label{eq:NL}
\phi^{\prime\prime}+2\frac{a^\prime}{a}\phi^\prime - \nabla^2\phi + a^2 V_{,\phi}=0,
\end{align}
where $\prime$ denotes derivative with respect to conformal time. This partial differential equation is solved by discretizing space on a grid of $N^3_{\rm pts}$ points and fixed comoving box size $L$, using periodic boundary conditions. Spatial derivatives are calculated using a second-order finite difference method. The time integration is done via a Runge-Kutta 4th order method, which solves at the same time the (second) Friedmann equation for the scale factor
\begin{equation}
a^{\prime\prime} = \frac{1}{6}\left(\langle\rho\rangle-3\langle p \rangle\right)a^3,
\end{equation}
where $\langle\rho\rangle$ and $\langle p \rangle$ are respectively the mean energy density and pressure contained in the lattice, computed as averages over the $N^3_{\rm pts}$ lattice points. In this setup, the metric is fixed to FLRW, neglecting the role of metric perturbations in the evolution of the inflaton field, which is justified as explained in Sec. \ref{sec:grav_back}.

\subsubsection{Initial conditions and discretization effects}
Although the simulation evolves the inflaton field nonlinearly, without separating the background from fluctuations, the initial conditions on the lattice are set perturbatively. The background inflaton value and its velocity are initialized to the homogeneous solution before the USR phase, while fluctuations are introduced in the discrete Bunch-Davies vacuum in Fourier space~\cite{Caravano:2021pgc,Caravano:2022yyv}:
\begin{align}
\label{eq:IC}
\delta\phi (\vec{\kappa},\tau)=\frac{1}{\sqrt{2{\omega}_{\vec\kappa}}}e^{-i \omega_{\vec\kappa} \tau},
\end{align}
where 
\begin{align}
    \vec\kappa=\frac{2\pi}{L}\{m_1,m_2,m_3\},\quad\quad m_i\in \{1,...,N_{\rm pts}\},
\end{align}
is the discrete lattice momentum, and 
\begin{align}
\label{eq:eff}
    \omega_{\vec\kappa} = \frac{2N_{\rm pts}}{L}\sqrt{\sum_{i=1}^3\sin^2\left(\frac{\pi m_i}{N_{\rm pts}}\right)}
\end{align}
is the effective momentum induced by the discrete lattice structure~\cite{Caravano:2021pgc}. This effective momentum $\omega_{\vec\kappa}$ converges to $|\vec\kappa|$ for $N_{\rm pts}\rightarrow \infty$. As common in lattice simulations, initial fluctuations on the lattice are then set starting from Eq. \eqref{eq:IC} via a stochastic process that mimics the quantum initial conditions (see~\cite{Caravano:2021pgc,Caravano:2022yyv} for details). 

The form of Eq. \eqref{eq:eff} is dictated by the choice of the stencil for the Laplacian operator in Eq. \eqref{eq:NL}, that we calculate using a second-order finite difference scheme. This momentum is identified with the physical one, $\omega_\kappa \leftrightarrow k$, when computing momentum-dependent observables from the simulation. This identification is crucial for obtaining power spectra from the simulation with enough precision. For further details on the calculation of the power spectrum, see~\cite{Caravano:2021pgc,Caravano:2022yyv}.

\subsubsection{Lattice resolution}
For this work, we use a lattice with $N^3_{\rm pts} = 512^3$ points. The box size $L$ is chosen such that:
\begin{align}
\label{eq:res}
   0.22 \,k_{\rm ref} \leq k \equiv \omega_{\kappa}\leq 45\, k_{\rm ref},
\end{align}
allowing us to capture the peak in the power spectrum. We start the simulation at $N=-2$, making all modes sub-Hubble at the beginning of the simulation. We end the simulation around $N=5$, when all modes are super-Hubble. 
See related discussion around Fig. \ref{fig:pzeta}. Although we show results for a box size with resolution given by eq.~\eqref{eq:res}, we run additional simulations with different resolutions, corresponding to changing lattice parameters $N_{\rm pts}$ and $L$, to ensure that our results are free of lattice artifacts, such as finite-size effects.

\subsection{Simulation results}

In this section we show simulation results for the different potentials introduced in Sec.~\ref{sec:setup}, namely cases I, II and III of Table \ref{tab:usr_cases}. For each of these cases, we run several lattice simulations for different potentials that are predicted to generate tree-level power spectra with a peak in the range $\mathcal{P}^{\rm max}_{\zeta,\rm tree}= 10^{-4}\div 1$.  In all plots presented in the following, each line from blue to yellow corresponds to a unit jump of $\log_{10}(\mathcal{P}^{\rm max}_{\zeta,\rm tree})$.

\subsubsection{Background dynamics}
\begin{figure*}
\centering

\centering
	
	\begin{tikzpicture}
  \node (img) at (-7,0) {\includegraphics[width=8.cm]{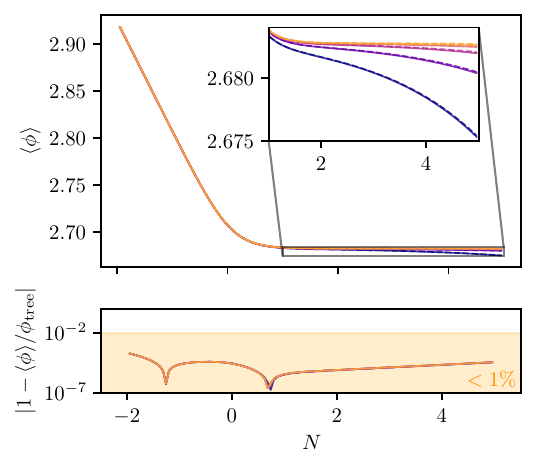}};
	\node (img2) at (1.3,0) {\includegraphics[width=8.cm]{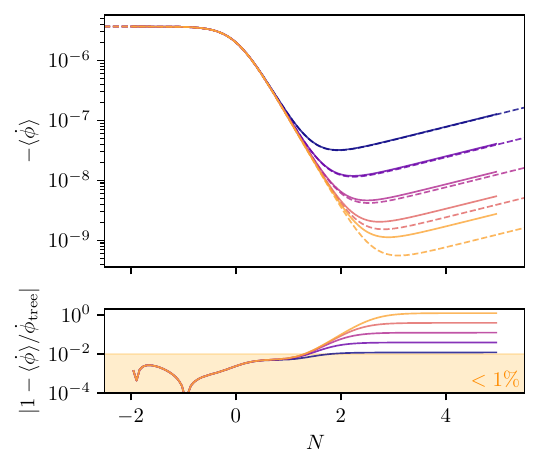}};

 \node (img) at (-7,0-7) {\includegraphics[width=8.cm]{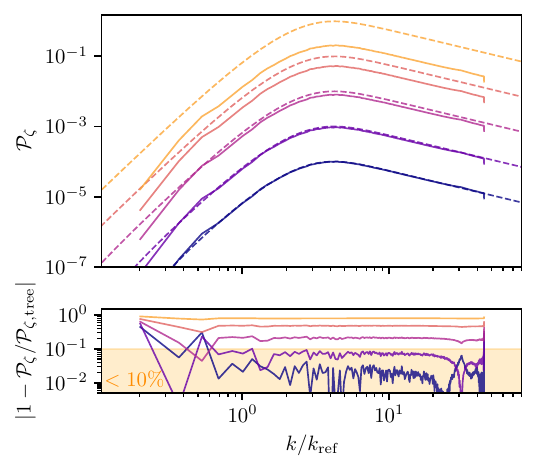}};
	\node (img2) at (1.3,0-7) {\includegraphics[width=8.cm]{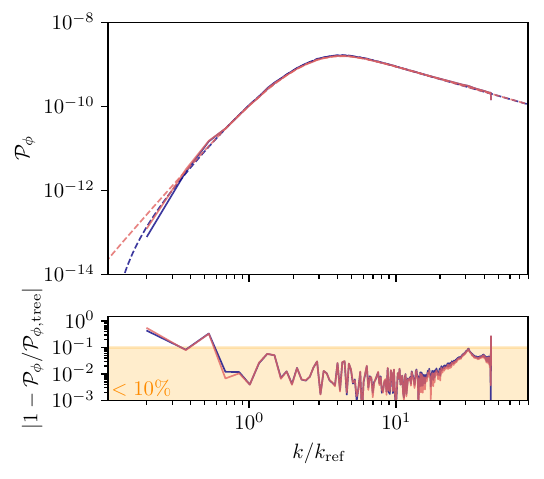}};

 \node (img) at (-7,0-14) {\includegraphics[width=8.cm]{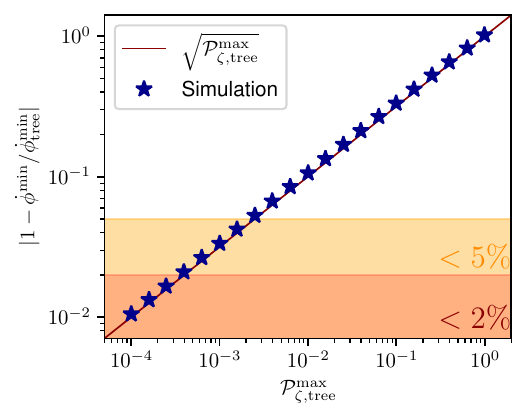}};
	\node (img2) at (1.3,0-14) {\includegraphics[width=8.cm]{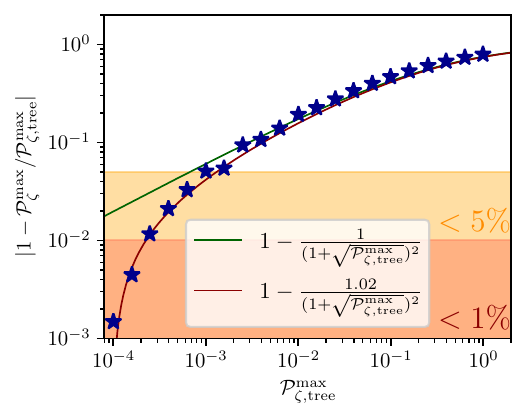}};

	\node [align=center] at (-2.2,4){\Large case I (Wands duality)};

	\end{tikzpicture}

\caption{ Simulation results for case I (Wands duality) of Table \ref{tab:usr_cases}. 
{\it Top panels:} 
Plot of background quantities $\langle\phi\rangle$ (left) and $\langle\dot\phi\rangle$ (right) computed as averages over the $N_{\rm pts}^3$ lattice points. The relative difference with respect to the purely homogeneous result is shown in the lower panel.
{\it Middle panels:} 
Power spectrum of $\zeta$ (left) and of $\phi$ (right) at the end of the simulation ($N=4.96$). Colors range from $\mathcal{P}^{\rm max}_{\zeta,\rm tree}=10^{-4}$ (blue) to $\mathcal{P}^{\rm max}_{\zeta,\rm tree}=1$ (yellow), similarly to Figs.~\ref{fig:dyn} and \ref{fig:dyn_pot}. 
Tree-level results from SPT are shown as dashed lines, with relative differences shown in the lower panel.
{\it Bottom panels:} 
Nonlinear correction to $\dot\phi$ (left panel) and to the power spectrum peak (right panel) as a function of the linear theory prediction. Each point of this plot is obtained from one lattice simulation.
}\label{fig:caseI}
\end{figure*}

\begin{figure*}
\centering

\centering
	
	\begin{tikzpicture}
  \node (img) at (-7,0) {\includegraphics[width=8.cm]{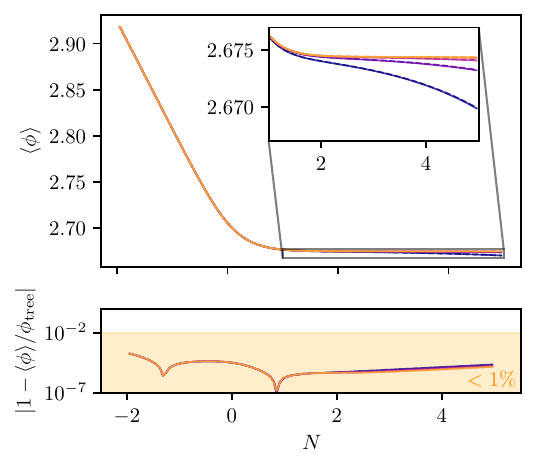}};
	\node (img2) at (1.3,0) {\includegraphics[width=8.cm]{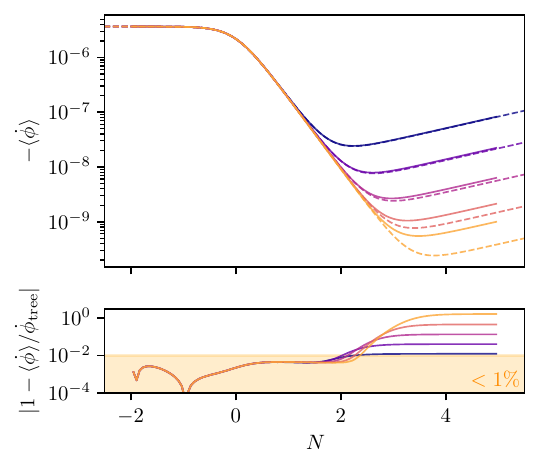}};

 \node (img) at (-7,0-7) {\includegraphics[width=8.cm]{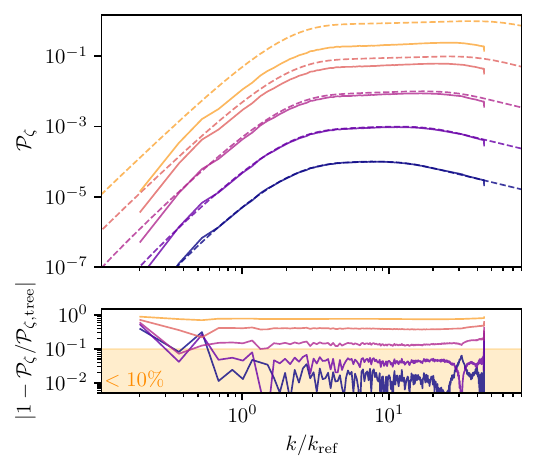}};
	\node (img2) at (1.3,0-7) {\includegraphics[width=8.cm]{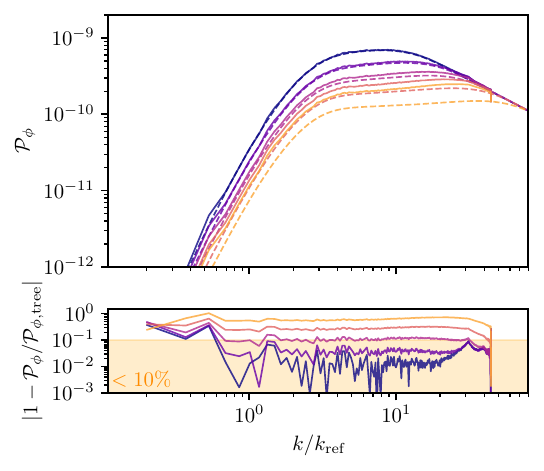}};

 \node (img) at (-7,0-14) {\includegraphics[width=8.cm]{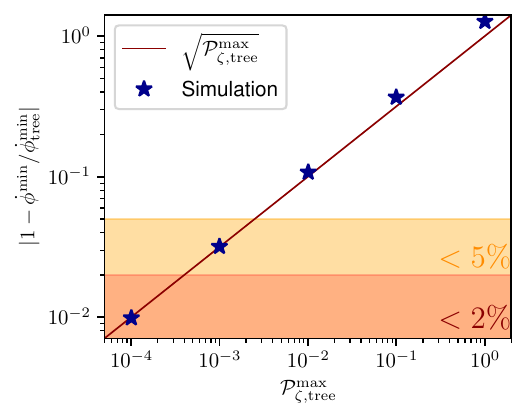}};
	\node (img2) at (1.3,0-14) {\includegraphics[width=8.cm]{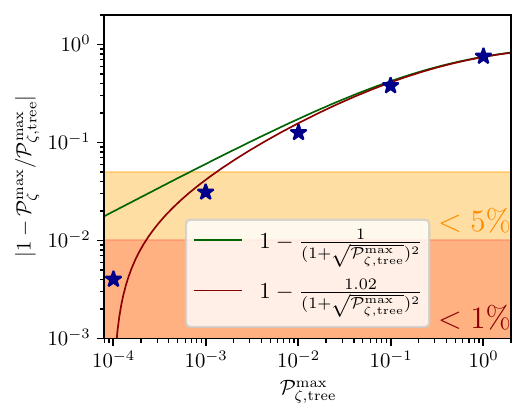}};

	\node [align=center] at (-2.2,4){\Large case II (repulsive)};

	\end{tikzpicture}

\caption{ Same as Fig. \ref{fig:caseI}, but for case II (repulsive). 
}\label{fig:caseII}
\end{figure*}

\begin{figure*}
\centering

\centering
	
	\begin{tikzpicture}
  \node (img) at (-7,0) {\includegraphics[width=8.cm]{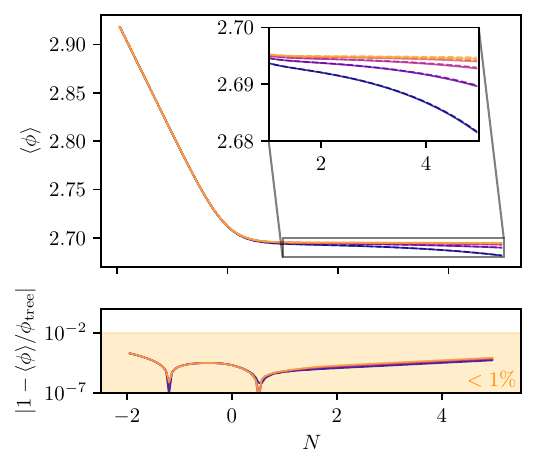}};
	\node (img2) at (1.3,0) {\includegraphics[width=8.cm]{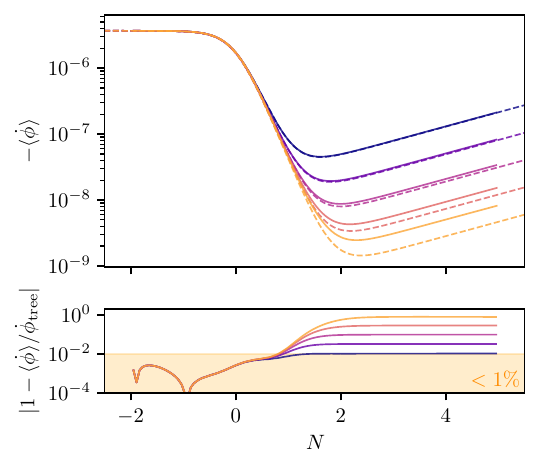}};

 \node (img) at (-7,0-7) {\includegraphics[width=8.cm]{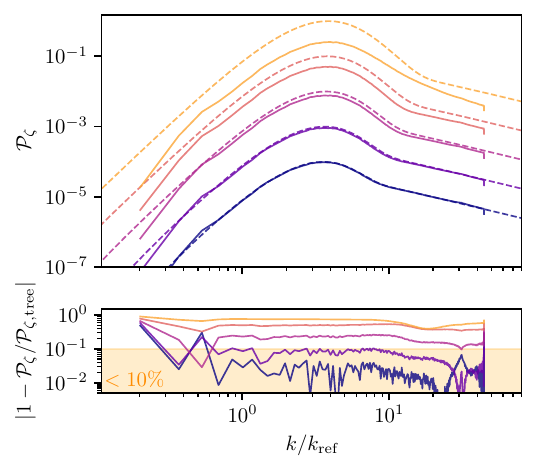}};
	\node (img2) at (1.3,0-7) {\includegraphics[width=8.cm]{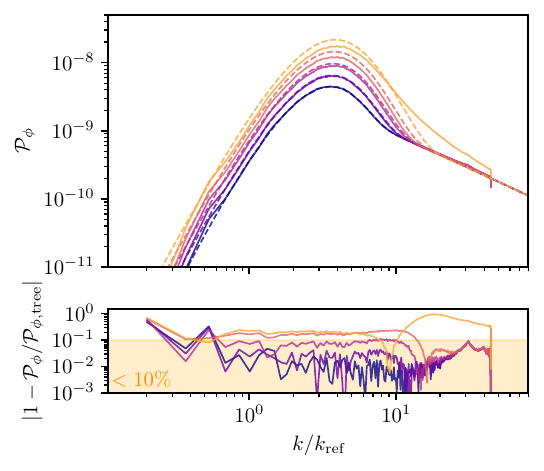}};

 \node (img) at (-7,0-14) {\includegraphics[width=8.cm]{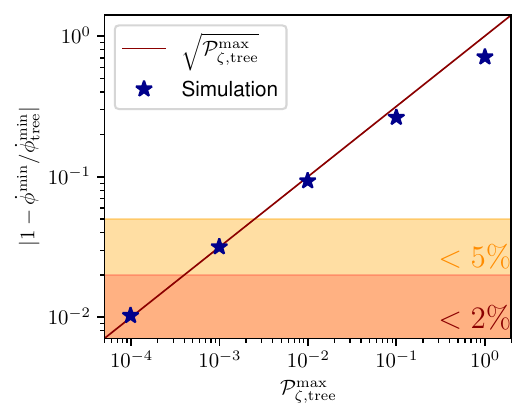}};
	\node (img2) at (1.3,0-14) {\includegraphics[width=8.cm]{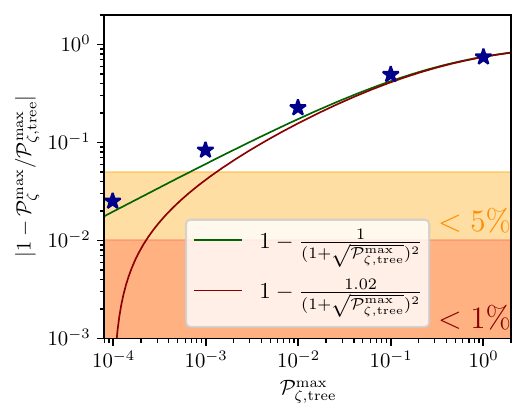}};

	\node [align=center] at (-2.2,4){\Large case III (attractive)};

	\end{tikzpicture}

\caption{ Same as Fig. \ref{fig:caseI}, but for case III (attractive).
}\label{fig:caseIII}
\end{figure*}

\begin{figure*}
\centering
\includegraphics[width=0.33\linewidth]{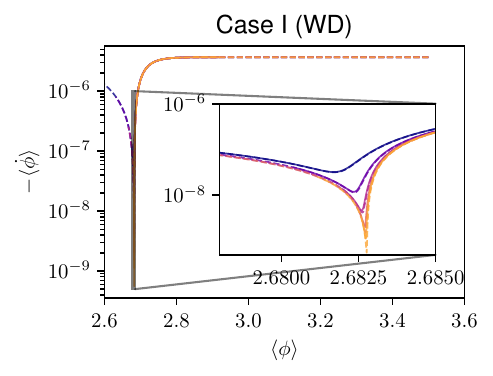}
\includegraphics[width=0.32\linewidth]{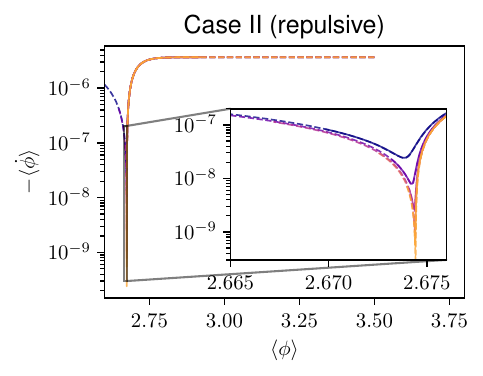}
\includegraphics[width=0.32\linewidth]{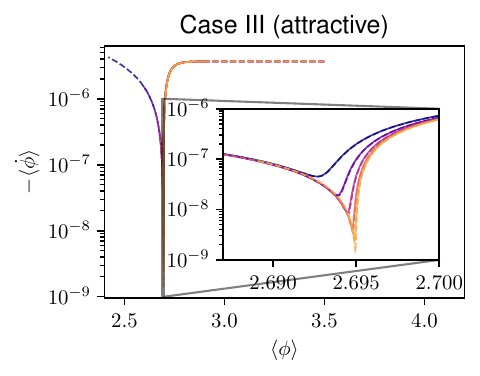}
\caption{Phase space trajectory $\dot\phi(\phi)$ from the simulation, compared with the purely homogeneous prediction (dashed lines). Different colors correspond to different $\mathcal{P}_{\zeta,\rm tree}^{\rm max}$ in the range $10^{-4 \div 0}$, like in all other plots. \label{fig:trajectory}}
\end{figure*}

First, we compare the background evolution with the solution of the homogeneous Klein-Gordon equation~\eqref{eq:EoM} that we obtain numerically. This comparison  is shown in the top panels of Figs. \ref{fig:caseI}, \ref{fig:caseII}, and \ref{fig:caseIII} for each of the three cases. In all cases, the background solution agrees with SPT at the $<1\%$ level for $\mathcal{P}^{\rm max}_{\zeta,\rm tree}= 10^{-4}$. However, for larger values of $\mathcal{P}^{\rm max}_{\zeta,\rm tree}$, the inflaton mean velocity $\langle\dot\phi\rangle$ in the lattice simulation deviates from SPT
predictions, with O(1) deviations for $\mathcal{P}^{\rm max}_{\zeta,\rm tree}=1$. 
In particular, during the USR phase, the velocity in the simulations does not decrease as much as predicted by the purely homogeneous evolution.
This is a consequence of the backreaction of fluctuations on the background evolution, which is not taken into account by SPT.
In Sec.~\ref{sec:intepretation}, we give tentative explanations of this phenomenon. Due to the very small inflaton velocity throughout the USR phase, the backreaction effect in the mean field value $\langle\phi\rangle$ as a function of number of $e$-folds is much less significant, but the two are naturally related, as the phase-space portrait in Fig.~\ref{fig:trajectory} will show explicitly. In particular, the field evolution returns to the SR attractor after the USR phase.

\subsubsection{Fluctuations}
 Next, we analyze the statistics of inflaton fluctuations $\delta\phi=\phi-\langle\phi\rangle$ and the comoving curvature perturbation $\zeta$, calculated from the lattice as $\zeta = -H \delta\phi /\dot\phi$, assuming a linear relation between $\delta\phi$ and $\zeta$. For large $\mathcal{P}^{\rm max}_{\zeta,\rm tree}$, this assumption may become inaccurate, but comparing lattice results with tree-level results of SPT (sometimes simply referred to as linear theory in the following) under this assumption is still useful to assess the validity of the perturbative picture. We will investigate the role of nonlinear relation between inflaton and curvature perturbations in future work.
 
 In the middle panels of Figs. \ref{fig:caseI}, \ref{fig:caseII}, and \ref{fig:caseIII}, we show the power spectra of $\zeta$ and $\phi$, and we compare them with the linear theory predictions. For $\mathcal{P}^{\rm max}_{\zeta,\rm tree}=10^{-4}$, both $\mathcal{P}_{\phi}$ and $\mathcal{P}_{\zeta}$ match linear predictions in all three cases. For larger $\mathcal{P}^{\rm max}_{\zeta,\rm tree}$, however, corrections appear, with O(1) deviations in $\mathcal{P}_{\zeta}$ for $\mathcal{P}^{\rm max}_{\zeta,\rm tree}\geq 0.1$, and deviations exceeding $\sim 20\%$ for $\mathcal{P}^{\rm max}_{\zeta,\rm tree}=10^{-2}$.\footnote{ One should be aware that the spectrum of perturbations at low $k$ in Figs. \ref{fig:caseI}, \ref{fig:caseII}, and \ref{fig:caseIII} are largely affected by sample variance due to the limited number of modes contained in the simulation box. }

The power spectrum of the inflaton $\mathcal{P}_{\phi}$ also shows significant deviations in cases II and III. Specifically, it is larger than the SPT prediction in case II and smaller in case III. In case I, however, it perfectly agrees with linear theory, giving the same $\mathcal{P}_{\phi}$ profile for all values of $\mathcal{P}^{\rm max}_{\zeta,\rm tree}$. This is a consequence of the weakly broken Wands duality characterizing this scenario, which results in the same approximately free theory for $\delta\phi$ in this case. In other words, in case I all interactions are nearly switched off, making the spectrum of perturbations insensitive to nonlinearities. Let us stress again, however, that a backreaction on the background evolution is still present in case I. 

\begin{figure*}
\includegraphics[width=18cm]{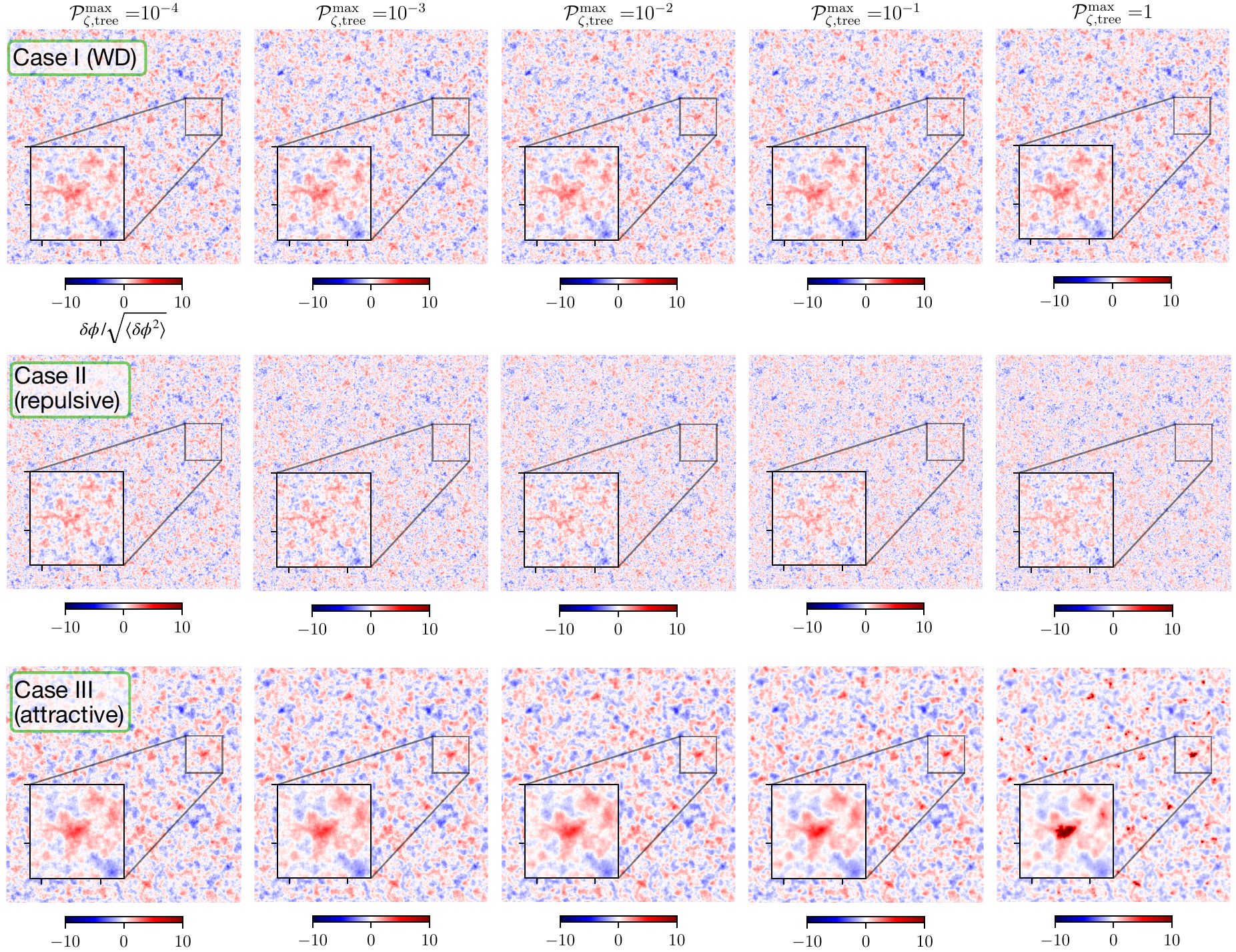}
\caption{2D snapshots of the inflaton field at the end of the simulation. Different columns correspond to different $\mathcal{P}^{\rm max}_{\zeta,\rm tree}$, while the rows refer to the three different cases. These simulations are run with the same initial conditions to  highlight the differences induced by the USR evolution between the three cases.}
\label{fig:snapshots}
\end{figure*}

It is instructive to compare the properties of the inflaton power spectrum with real-space snapshots of the inflaton field at the end of the simulation, shown in Fig. \ref{fig:snapshots}. These snapshots are normalized by the variance of the inflaton field, making them insensitive to the amplitude of fluctuations. This helps to illustrate the structural differences in the inflaton field across cases and their connection to statistical properties:
\begin{itemize}[leftmargin=3mm]

\item In case I, structures in the inflaton field remain unchanged, confirming the nearly-free nature due to the approximate Wands duality. 

\item In Case II, the granularity of structures increases as nonlinear evolution transfers power to small scales, consistent with the middle-right panel of Fig.~\ref{fig:caseII}, where we see that the small-scale power spectrum grows with larger $\mathcal P_{\zeta,\rm tree}^{\rm max}$.
This is also partially due to the broader spectrum of perturbations, enhancing higher $k/k_{\rm ref}$ with respect to the other cases, due to the choice of $\eta_{\rm II} = 3$. Furthermore, one sees that ``red structures'' corresponding to positive $\delta \phi$, like the one we have zoomed in, are less red in case II than in case I, which one can attribute to the repulsive self-interactions in this scenario. Indeed, with a positive cubic coupling $V_3$, configurations with $\delta \phi >0$ are more costly energetically than in the free theory. This effect also explains why red regions smooth out when increasing  $\mathcal P_{\zeta,\rm tree}^{\rm max}$: as is clear from Fig.~\ref{fig:dyn_pot}, $V''(\phi)$ varies more sharply as $\mathcal P_{\zeta,\rm tree}^{\rm max}$ increases, corresponding to a larger strength of the cubic coupling $V_3$, and hence larger repulsive effects.

\item In case III, the typical size of the structures remains the same as $\mathcal P_{\zeta,\rm tree}^{\rm max}$ is enhanced. This is consistent with the middle-right panel of Fig. \ref{fig:caseIII}, where the position of the sharp peak in the power spectrum is not affected by nonlinearity (only its amplitude). Moreover, contrary to case II, red regions are redder here than in the case of Wands duality: with a negative cubic coupling $V_3$, configurations with positive $\delta \phi$ are energetically favored compared to the free theory, i.e.~the attractive self-interaction induces more clustering. This also explains why red regions become even more red as $\mathcal P_{\zeta,\rm tree}^{\rm max}$ increases: this corresponds to sharper variations of $V''(\phi)$, and hence to a larger magnitude of the attractive coupling.  

\end{itemize}

Let us stress that  while these effects are most evident in the extreme cases where the spectrum of perturbation is large, they are already visible for moderately large enhancements. This supports the expectation of important effects from nonlinear physics in the phenomenology of USR scenarios already for $\mathcal{P}^{\rm max}_{\zeta} \sim {\cal O}(10^{-2})$.

\begin{figure*}
\centering
\includegraphics[ width=0.34 \linewidth]{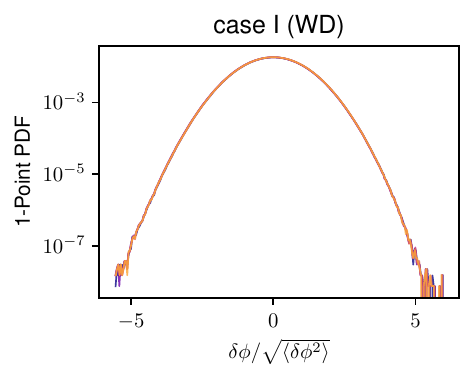}
\includegraphics[width=0.32 \linewidth]{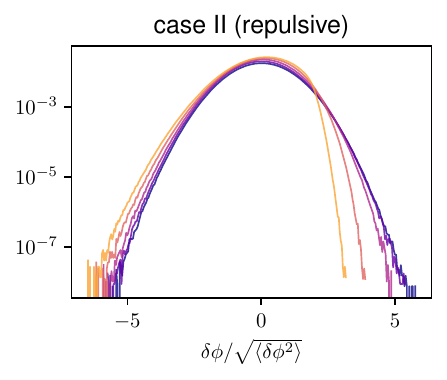}
\includegraphics[width=0.32 \linewidth]{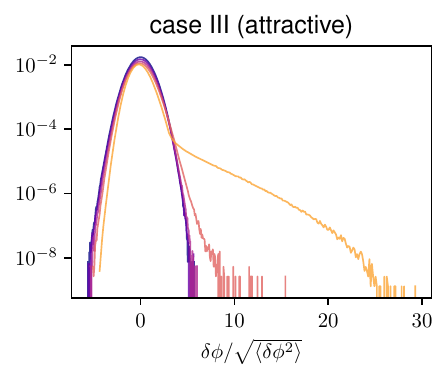}
\caption{1-point PDF of $\delta\phi=\phi-\langle\phi\rangle$ at the final time $N=4.96$. Colors correspond to different peak values of $\mathcal{P}_{\zeta,\rm tree}^{\rm max}$ in the range $10^{-4 \div 0}$. As we normalize by the standard deviation $\sqrt{\langle\delta\phi^2\rangle}$, any deviation from the nearly Gaussian case $\mathcal{P}_{\zeta,\rm tree}^{\rm max} = 10^{-4}$ (blue) represents a deviation from Gaussianity.}
\label{fig:PDF}
\end{figure*}

Eventually, it is also interesting to look at the 1-point probability density function (PDF) of the inflaton fluctuation at the final simulation time, shown in Fig. \ref{fig:PDF} for all three cases. Contrary to the snapshots, this does not carry any morphological information, but it gives a straightforward visualization of some statistical properties.
Like in Fig.~\ref{fig:snapshots}, we plot the distribution with the $x-$axis normalised to the perturbation variance $\sqrt{\langle \delta \phi^2 \rangle}$. Therefore, any deviation from the blue curve, for which nonlinear effects are minimal and negligible, indicate non-Gaussianity of the distribution. 
In case I, the statistics remain nearly Gaussian for all values of $\mathcal{P}^{\rm max}_{\zeta,\rm tree}$, consistent with the nearly-free nature of this model at the relevant times of the dynamics. 
In cases II and III, non-Gaussianity grows with increasing $\mathcal{P}^{\rm max}_{\zeta,\rm tree}$. 
In case II, this manifests as an increasingly negatively skewed distribution, a suppression of the right tail with $\delta \phi >0$, and a (more modest) enhancement of the left tail with $\delta \phi <0$. This behavior of the tails qualitatively agrees with our discussion on the repulsive interaction, but we stress that the latter aspect is better identified with the morphological information visible in the snapshots, and that the tails, by definition, are also sensitive to higher-order couplings. 
Similar remarks hold for case III with inverse trends: the larger $\mathcal{P}^{\rm max}_{\zeta,\rm tree}$, the more positively skewed the distribution, the smaller the left-tail and the larger the right tail. We highlight that the striking behavior of the right-tail in the orange most extreme scenario (with $\mathcal{P}^{\rm max}_{\zeta,\rm tree}=1$) is a manifestation of a phenomenon that is qualitatively different than the other situations. This tail reflects that, while most patches manage to escape the USR phase, some lack sufficient kinetic energy to climb the potential maximum and instead roll back toward the minimum. This is illustrated by the animation of the simulations (\href{https://github.com/caravangelo/USR-on-the-lattice.git}{link}), where we show the inflaton 1-point PDF moving along the potential. These patches, trapped in a false vacuum, will form PBH after inflation, as described in \cite{Caravano:2024tlp,Inomata:2021tpx,Atal:2019cdz,Atal:2019erb,Escriva:2023uko,Flores:2024lng}. {By contrast, for $\mathcal{P}^{\rm max}_{\zeta,\rm tree}=0.1$, the exponential tail reflects intrinsic non-Gaussianity of the inflaton field rather than the trapping phenomenon, except for a few statistically insignificant extreme points.}

\subsubsection{Discussion}
\label{sec:intepretation}
We now discuss some physical interpretation of the lattice results.

In case I, the spectrum of $\delta\phi$ remains unaffected by nonlinear effects due to the near-free nature of this scenario. Nonlinearities here arise entirely as a backreaction of fluctuations on the inflaton's background velocity. In the left-bottom corner of Fig. \ref{fig:caseI}, we show the nonlinear correction to the minimum value reached by $|\dot\phi|$, corresponding to the end of the USR phase, for different values of $\mathcal{P}^{\rm max}_{\zeta,\rm tree}$. We find the following relation between the fully nonlinear inflaton velocity and the one obtained from linear theory at the end of the USR phase
\begin{align}
\label{eq:corr_phi}
\langle \dot\phi \rangle=\dot\phi_{\rm tree}\left(1+\sqrt{\mathcal{P}^{\rm max}_{\zeta,\rm tree}}\right).
\end{align}

Plugging this into the linear relation $\zeta= -H \delta\phi/\dot\phi$, and noting that $\delta\phi$ is unaffected by nonlinearity, we obtain:
\begin{align}
\label{eq:corr_pzeta}\mathcal{P}_{\zeta} = \frac{\mathcal{P}_{\zeta,\rm tree}}{\left(1+\sqrt{\mathcal{P}^{\rm max}_{\zeta,\rm tree}}\right)^2}.
\end{align}
In the bottom-right corner of Fig. \ref{fig:caseI}, we compare lattice results with this expression, using the maximum $\mathcal{P}_{\zeta}$ values from various simulations. The green curve represents this relation, which holds in the regime of significant nonlinear effects ($>5\%$), with deviations below $2\%$, possibly due to small simulation systematics or other minor effects.

In cases II and III, nonlinearity also affects inflaton fluctuations, as we already discussed in the previous section. Despite that, in the bottom panels of Figs. \ref{fig:caseII} and \ref{fig:caseIII}, we can see that the nonlinear relations 
\eqref{eq:corr_phi} and \eqref{eq:corr_pzeta} are approximately satisfied also in these cases. This suggests that despite nonlinear dynamics in $\delta\phi$, the dominant nonlinear effect on $\mathcal{P}_{\zeta}$ still comes from backreaction on $\dot\phi$. Nevertheless, the phenomenology of these models, in particular in relation to the probability of PBH formation, will be highly sensitive to the non-Gaussian features discussed above, making nonlinear corrections to $\delta\phi$ statistics crucial in these cases.

\begin{figure*}
\centering
\includegraphics[width=5.8cm]{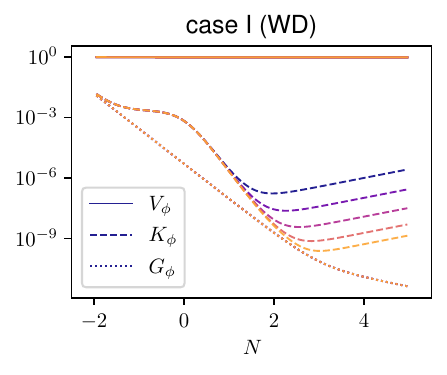}
\includegraphics[width=5.8cm]{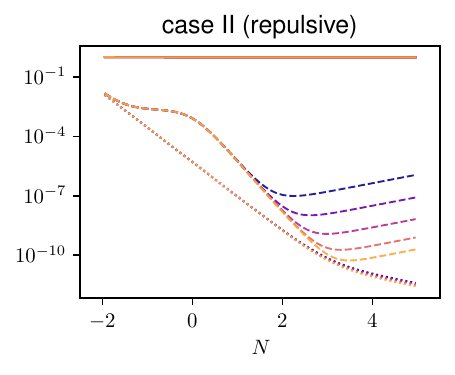}
\includegraphics[width=5.9cm]{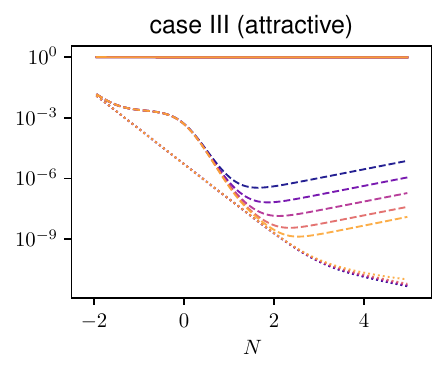}
\caption{Relative contributions to the energy density in the simulation as a function of $e$-folds time $N$. Full lines represent the potential energy $V_{\phi}$, while dashed lines and dotted line are respectively the kinetic ($K_{\phi}$) and gradient energy ($G_{\phi}$) in the simulation. Different colors correspond to different $\mathcal{P}_{\zeta,\rm tree}^{\rm max}$ in the range $10^{-4 \div 0}$. \label{fig:energy}}
\end{figure*}

To better understand backreaction in this system, let us consider some analytical arguments. By definition, the SPT background, denoted $\phi_{\rm st}$ in the following, satisfies the homogeneous Klein-Gordon equation:
\begin{align}
\label{eq:KG}
\ddot\phi_{\rm st}+3H_{\rm st}\dot\phi_{\rm st}+V^\prime(\phi_{\rm st})=0.
\end{align}
The “true” simulation background is found by averaging \eqref{eq:NL}, yielding
\begin{align}
\label{eq:KG_lat}
\langle\ddot\phi\rangle+3H\langle\dot\phi\rangle+\langle V^\prime(\phi)\rangle=0.
\end{align}
Expanding the potential around $\phi_{\rm st}$:
\begin{align}
\begin{split}
V^\prime(\phi)=V^\prime(\phi_{\rm st})+&\left(\phi-\phi_{\rm st}\right)V^{\prime\prime}(\phi_{\rm st})\,+\\&+\frac{\left(\phi-\phi_{\rm st}\right)^2}{2}V^{(3)}(\phi_{\rm st})+...\,,
\end{split}
\end{align}
defining $\delta \phi_B = \langle\phi\rangle-\phi_{\rm st}$ and subtracting Eq.~\eqref{eq:KG} from \eqref{eq:KG_lat} gives:
\begin{align}
\ddot{\delta \phi_B}+3H\dot{ \delta \phi_B}+V^{\prime\prime}(\phi_{\rm st})\delta \phi_B \simeq - \frac{\langle\left(\phi-\phi_{\rm st}\right)^2\rangle}{2}V^{(3)}(\phi_{\rm st}),
\label{eq:deltaphiB}
\end{align}
where we assumed $H_{\rm st} = H$, as verified by our lattice simulations.
This equation indicates that even in the WD case, where field interactions become negligible deep in the USR phase (rendering the right-hand side negligible), any small difference $\delta \phi_B$ will be exponentially amplified by the large tachyonic mass during USR, regardless of its sign or the specific nonlinear mechanism that initially produced $\delta \phi_B$.

We now present a complementary potential physical interpretation of the backreaction on the inflaton velocity. The potentials considered here have a field value where the velocity is minimized, typically reached around $N \sim 2$. In cases I and III, this corresponds to a local maximum in the potential. Backreaction in the inflaton velocity arises when the mean field is near this point, corresponding to the end of the feature in the potential. At this stage, the inflaton is distributed across various values, with a variance determined by the power spectrum. This distribution behaves non-uniformly: some regions that advance the mean proceed more quickly toward the second slow-roll attractor. Without gradients, these faster-moving regions would evolve independently; however, they are linked to slower regions through gradient terms, which exert a ``pull'' that accelerates the slower regions via gradient forces, resulting in an overall acceleration of the background. This mechanism is consistent with the animations from our lattice simulations, available at this \href{https://github.com/caravangelo/USR-on-the-lattice.git}{link}. In these animations, we show snapshots of the simulation together with the 1-point PDF of the inflaton field (blue) moving on top of the potential (green), and compare the simulation average value $\langle\phi\rangle$ (red) with the homogeneous solution $\phi_{\rm st}$ (dashed red), which neglects backreaction. The animations are for $\mathcal{P}^{\rm max}_{\zeta, \rm tree}=1$. The animations show that backreaction (i.e. the difference between the red line and the dashed red line) arises precisely when the system reaches the point of minimum velocity.
This supports our tentative interpretation, as by this time some regions are already drifting towards the second SR attractor. However, note that nonlinear corrections to the potential can independently induce backreaction. Therefore, the observed backreaction may arise from multiple contributing factors.

At this stage, two remarks are important. First, the effects of gradients discussed above are not taken into account in the separate universe picture.
Second, if this interpretation is correct, the additional acceleration of the inflaton trajectory is temporary and ceases once most points have crossed the USR phase (consistent with Eq.~\eqref{eq:deltaphiB} once the $V''$ and $V^{(3)}$ terms have become negligible). Thus, backreaction should manifest as an advancement in the background motion, but the background trajectory $\dot\phi(\phi)$ should fall back to the same slow-roll attractor in the second SR phase, as already suggested by results in the top right plots of Figs. \ref{fig:caseI}-\ref{fig:caseIII}. This is confirmed by the simulations, as we present in Fig. \ref{fig:trajectory}, where we show that the phase space trajectory $\dot\phi(\phi)$ coincides with the homogeneous prediction of SPT without backreaction sufficiently before and after the USR phase, in all cases considered.

Finally, let us discuss the backreaction on $\dot\phi$ in relation to the evolution of different energy contributions during the USR phase. In Fig. \ref{fig:energy}, we show the evolution of the different relative contributions to the energy density in the lattice simulation for the three different cases. When linear perturbation theory holds (e.g., $\mathcal{P}_{\zeta}=10^{-4}$), there is a clear hierarchy among the energy contributions, $V_\phi \gg K_\phi \gg G_\phi$, similar to standard slow-roll inflation. 
Backreaction becomes significant when the kinetic energy decreases to the level of the gradient energy, $K_\phi \sim G_\phi$, which is not taken into account in SPT.\footnote{Note that $K_{\phi}\sim G_{\phi}$ also at the starting time $N=-2$, when the simulation box is sub-Hubble. This is common in lattice simulations where one initializes sub-Hubble quantum fluctuations as stochastic noise. Contrary to what happens at $N\simeq 2$, this does not affect the dynamics of the field, which on sub-Hubble scales is trivial and is dominated by the gradient term in the equation of motion. Therefore, at the initial time, the interplay between gradients and other energy contributions is irrelevant. This is confirmed by the agreement between the lattice simulation and linear theory at early times.} Moreover, we have seen that the kinetic energy is larger than the one predicted by SPT, suggesting that backreaction effectively protects the kinetic energy from dropping below the gradient energy density.

\section{Conclusions and outlook}\label{sec:conclusions}

In this work, we studied the nonlinear dynamics and backreaction effects in inflationary scenarios featuring a transient USR phase using lattice simulations. We demonstrated that for power spectra peaking at \( \mathcal P_{\zeta}^{\rm max} \sim 10^{-2} \), nonlinear effects introduce {$\sim 20$\% }corrections to the power spectral amplitude, consistent with the order-of-magnitude expectation from perturbation theory {of $O(10\%)$}. {Accurately accounting for these corrections is essential to make reliable predictions for the generation of PBHs and SIGWs. Interestingly, the size of these effects suggests that perturbativity is not completely violated (in contrast to what happens, for example, in \cite{Caravano:2024tlp}). This indicates that perturbative computations beyond tree level—when incorporating the key backreaction effects identified here—could provide a viable analytical description. These effects would appear as large loop corrections around the SPT background, which should be resummed into a backreacted background around which
perturbation theory should be performed.}
%is performed.}
%\textcolor{blue}{Accurately accounting for these nonlinear corrections is essential for making reliable predictions for the generation of PBHs and SIGWs.
%In particular, for large amplifications of the power spectrum, what would appear as large loop corrections around the background of standard perturbation theory is partially resummed into a backreacted background that considerably differs from the former, and around which perturbation theory should be performed, offering interesting prospects for analytical understanding.}

%We showed that, already for power spectra peaking at \( \mathcal P_{\zeta}^{\rm max} \sim 10^{-2} \), nonlinear effects lead to corrections of the order of 10\% on the power spectral amplitude, in agreement with naive order of magnitude expectation of perturbation theory. 
%Importantly though, these corrections occur without violating the conditions for perturbativity, suggesting analytical perturbative computations beyond tree level could be sufficient to describe these models, once the important backreaction effect pointed out in this work is taken into account.   
%As USR inflationary models are particularly relevant for scenarios predicting the formation of PBHs and SIGWs, controlling 
%these nonlinear corrections is crucial for making robust predictions.

We have also shown that in scenarios characterized by an approximate Wands duality, nonlinear self-interactions remain suppressed, and the system behaves nearly as a free theory. In these cases, while backreaction effects on the background evolution are still present, the spectrum of scalar perturbations remains unaffected by nonlinear corrections, as previously suggested in the literature.
However, for models that break Wands duality, we observed more pronounced nonlinear interactions, leading to significant non-Gaussian features in the statistics of the inflaton perturbations. In particular, we found that for repulsive cases (characterized by a positive cubic interaction), the distribution of field fluctuations exhibits a suppression of large positive fluctuations, while for attractive cases (with negative cubic interactions), the field develops a heavy exponential tail. This behavior is critical for understanding the potential formation of PBHs, as the tail of the distribution directly affects the probability of generating large overdensities that could collapse into black holes. 
{Crucially, the exponential tail observed in this work reflects the intrinsic non-Gaussianity of the inflaton field and not that of $\zeta$. Therefore, this tail cannot be directly compared to those arising from the nonlinear relation between $\delta\phi$ and $\zeta$ or from stochastic effects (see, e.g., \cite{Cai:2018dkf,Ezquiaga:2019ftu}). Note that, in contrast to our method, these approaches neglect the intrinsic non-Gaussianity prior to the crossing of the coarse-graining scale.}

It was recently suggested that USR dynamics advocated by some of the PBH formation scenarios may violate perturbativity and induce loop corrections affecting much longer modes associated with the scales observed through the CMB \cite{Kristiano:2022maq}.\footnote{Instead, see \cite{Fumagalli:2023loc} for effects at near infrared scales when the power spectrum enhancement is due to particle production.} {While the existence and magnitude of this effect are still under debate \cite{Cheng:2021lif,Riotto:2023hoz,Kristiano:2023scm,Riotto:2023gpm,Kristiano:2024vst,Franciolini:2023lgy, Davies:2023hhn,Firouzjahi:2023ahg, Iacconi:2023ggt,Motohashi:2023syh,Tasinato:2023ioq, Tasinato:2023ukp,Firouzjahi:2023aum,Fumagalli:2023hpa, Cheng:2023ikq,Tada:2023rgp,Firouzjahi:2023bkt,Braglia:2024zsl, Kawaguchi:2024lsw,Ballesteros:2024zdp,Inomata:2024lud,Fumagalli:2024jzz,Kawaguchi:2024rsv,Green:2024fsz}, we can not directly address this question due to the relatively small range of scales that can be tracked in the lattice box. However, we demonstrated the existence of significant backreaction on the background dynamics, typically captured by tadpole contributions, which have been suggested to play a crucial role in protecting long modes from loop corrections \cite{Inomata:2024lud,Fumagalli:2024jzz}.}

Looking forward, it would be interesting to extend the present work in different directions. First, it will be important to derive the full non-Gaussian statistics for the curvature perturbation, including the nonlinear relation between $\zeta$ and $\delta \phi$ from our lattice simulations. This would allow us to assess, for example, whether positive non-Gaussianities in $\zeta$ are obtained in all scenarios considered here, as suggested in~\cite{Firouzjahi:2023xke}.
Other groups attempted to study the effect of backreaction and nonlinearities in the USR scenarios employing full \textit{in}-\textit{in} computations, $\delta N$, or the stochastic inflation formalism (see, e.g., \cite{Pattison:2017mbe, Biagetti:2018pjj, Ezquiaga:2019ftu, Ballesteros:2020sre, Pattison:2021oen, Cai:2022erk,Biagetti:2021eep,Figueroa:2021zah,Raatikainen:2023bzk,Li:2023zva,Mizuguchi:2024kbl,Ballesteros:2024zdp,Animali:2024jiz,Jackson:2024aoo}), and it would be interesting to compare these techniques with lattice results. 
Additionally, evaluating the four-point function of the field fluctuations $\delta \phi$ would provide key insights into the non-perturbative source of SIGW (complementing recent efforts in this direction \cite{Cai:2018dig,Unal:2018yaa,Yuan:2020iwf,Atal:2021jyo,Adshead:2021hnm,Abe:2022xur,Chang:2022nzu,Garcia-Saenz:2022tzu,Li:2023qua,Perna:2024ehx,Ruiz:2024weh}). 
These next steps will allow us to better quantify the importance of nonlinearities in the USR scenarios, paving the way for robust observational predictions for GW experiments, as well as constraints on the PBH abundance.

\let\oldaddcontentsline\addcontentsline% Store \addcontentsline
\renewcommand{\addcontentsline}[3]{}% Make \addcontentsline a no-op
\begin{acknowledgments}
We especially thank Sebastian Zell and Guillermo Ballesteros for proposing to simulate USR prior to this work and for interesting discussions. We also thank Keisuke Inomata, Yuichiro Tada and Vincent Vennin for insightful discussions. A.C. acknowledges funding support from the Initiative Physique des Infinis (IPI), a research training program of the Idex SUPER at Sorbonne Universit\'e. During the course of this work, S.RP was supported by the European Research Council under the European Union’s Horizon 2020 research and innovation programme (grant agreement No
758792, Starting Grant project GEODESI). This article is distributed under the Creative Commons Attribution International Licence
(\href{https://creativecommons.org/licenses/by/4.0/}{CC-BY 4.0}).
\end{acknowledgments}

\twocolumngrid
\bibliography{main}
\let\addcontentsline\oldaddcontentsline% Restore \addcontentsline

\end{document}